\documentclass[11pt,twoside,reqno]{amsart}
\ifx\ProceedingsGelomur\relax\else

\usepackage{xy,graphics}

\xyoption{all}
\def \Hil {\mathcal{H}}

\def \F {\mathcal{F}}

\def  \S{\mathcal{S}}

\def \R {\mathbb{R}}

\def \C {\mathbb{C}}
\def\u{\mathfrak{u}}

\def \Tr {\mathrm{Tr}}
\def\uno{\mathbb{I}}
\def\pd#1#2{\frac{\partial #1}{\partial #2}}

\font\frak=eufm10 scaled\magstep1

\font\bigblack=msbm10 scaled\magstep 2
\font\bbigblack=msbm10 scaled\magstep3

\def\goth #1{\hbox{{\frak #1}}}

\def\bigfield #1{\hbox{{\bigblack #1}}}
\def\bbigfield #1{\hbox{{\bbigblack #1}}}

\def\v #1{\vert #1\vert}             
 
\def\m #1 #2{(-1)^{{\v #1} {\v #2}}} 
\def\pd#1#2{\frac{\partial#1}{\partial#2}}

\def\<#1>{\langle#1\rangle}        
\def\>#1{{\bf #1}}                
\def\f(#1,#2){\frac{#1}{#2}}

\def\braket#1#2{\langle#1\mathbin\vert#2\rangle} 

\def\dt2#1{\frac{d^2 #1}{dt^2}}

\def\uno{\relax{\rm 1\kern-.28 em I}}

\def\F{{\sp F}}

\def\R{{\hbox{{\field R}}}}             
\def\big R{{\hbox{{\bigfield R}}}}
\def\bbig R{{\hbox{{\bbigfield R}}}}
\def\C{{\hbox{{\field C}}}}         

\def\dim{\hbox{{\rm dim}}}        
 
\def\Tr{\hbox{{\rm Tr}}} 

\def\ad{{\hbox{ad}}}

\newtheorem{theorem}{Theorem}

\newtheorem{proposition}{Proposition}
\newtheorem{definition}{Definition}
\newtheorem{lemma}{Lemma}
\newtheorem{example}{Example}
\newtheorem{remark}{Remark}

\def\R{\Bbb R}
\def\C{\Bbb C}

\def\Hil{{\cal H}}

\font\frak=eufm10 scaled\magstep1

\def\goth #1{\hbox{{\frak #1}}}
\def\<#1>{\langle#1\rangle}

\def\braket#1#2{\langle#1|#2\rangle}

\def \Hil {\mathcal{H}}
\def \F {\mathcal{F}}
\def \u{\mathfrak{u}}
\def \S{\mathcal{S}}
\def \O{\mathcal{O}}

\def \Tr {\mathrm{Tr}}
\def \R {\mathbb{R}}
\def \C {\mathbb{C}}

\begin{document}

\title[Introduction to Quantum Mechanics]{Introduction to Quantum Mechanics
  and the Quantum-Classical transition}%
\author{Jos\'e F. Cari\~nena}
\address{Departamento de F\'{\i}sica Te\'orica \\Universidad de Zaragoza \\
Ciudad Universitaria \\
50009 Zaragoza (SPAIN)}
\author{Jes\'us Clemente-Gallardo}

\address{BIFI-Universidad de Zaragoza \\
Corona de Arag\'on 42 \\
50009 Zaragoza (SPAIN)}
\author{Giuseppe Marmo}
\address{Dipartamento de Scienze Fisiche \\
Universit\'a Federico II and INFN Naples \\
Via Cintia I \\
80126 Naples (ITALY)}
\date{}%
\maketitle

\begin{abstract}
  In this paper we present a survey of the use of differential geometric
  formalisms to describe Quantum Mechanics. We analyze Schr\"odinger and
  Heisenberg frameworks from this perspective and discuss how the momentum map
  associated to the action of the unitary group on the Hilbert space allows to
  relate both approaches. We also study Weyl-Wigner approach to Quantum
  Mechanics and discuss the implications of bi-Hamiltonian structures at the
  quantum level.
\end{abstract}


\section{Introduction}

\subsection{The need for a quantum theory and relevant mathematical 
structures}

Interference phenomena of material particles (say, electrons, neutrons, etc)
provide us with the most convincing evidence for the need to elaborate a new
mechanics which goes beyond and encompasses classical mechanics. At the same
time, `corpuscular' behaviour of radiation, light, as exhibited in phenomena
like photoelectric and Compton effects shows that also the description of
radiation has to undergo deep changes. The relation between the corpuscular-like
and the wave-like behaviour is fully captured by the following equation that we
may call the {\bf Einstein--de Broglie} relation
\begin{equation}
  \label{eq:einsteindebroglie}
  p_j\,dx^j-E\,dt=\hbar (k_j\,dx^j-\omega\, dt).
\end{equation}

This relation between the Poincar\'e 1-form on the phase-space over space-time
and the optical 1-form on the optical phase-space establishes a relation
between momentum and energy of the `corpuscular' behaviour and the frequency
of the `wave' behaviour. The proportionality coefficient is the Planck
constant.

The way we use this relation is to predict under which experimental conditions
light of a given wave length and frequency would be detected as a corpuscle
with a corresponding momentum and energy and vice-versa (i.e. when an electron
would be detected as a wave in the appropriate experimental conditions).

If we examine more closely an interference experiment, like the double slit
one, we find some peculiar aspects for which we do not have a simple
interpretation in the classical setting.

If we perform the experiment in such a way that we make sure that,
at each time, only one electron is present between the source and the screen,
we find that the electron impinges on the screen at `given points'. 

\begin{center}
\resizebox{!}{4cm}{\includegraphics{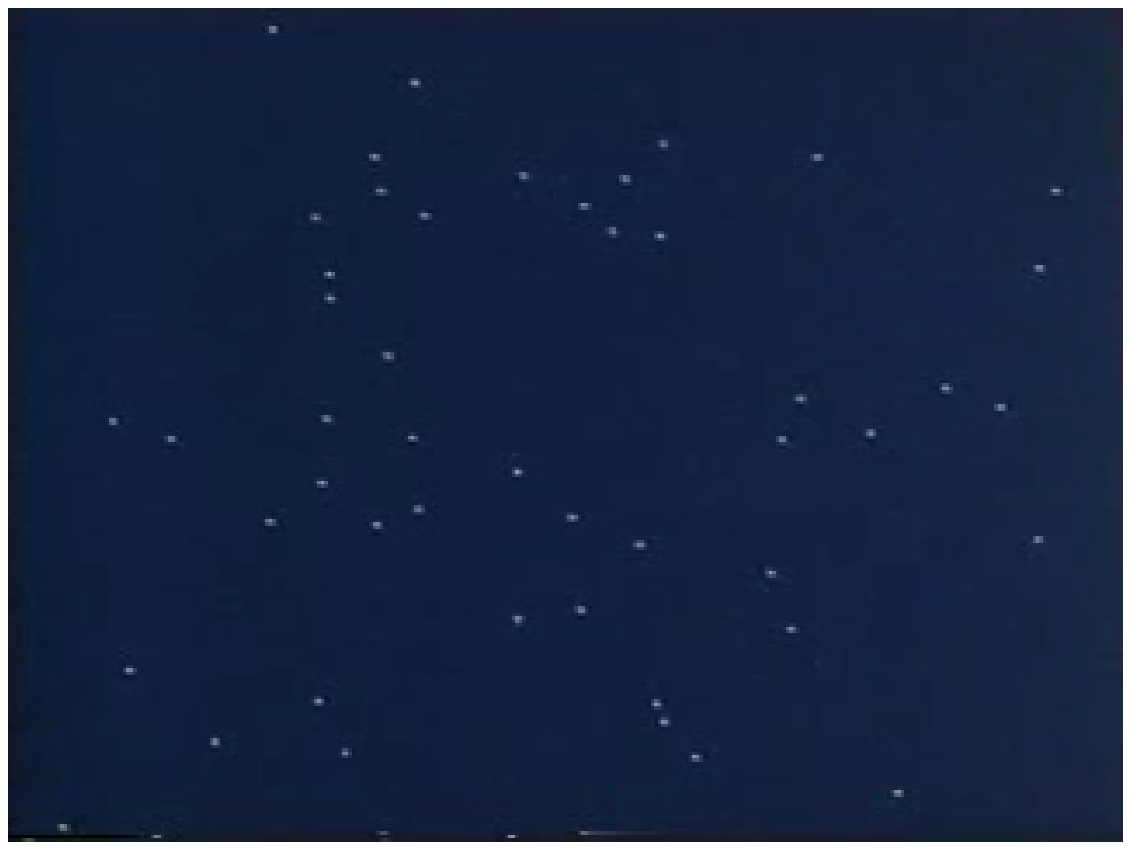}}
\resizebox{!}{4cm}{\includegraphics{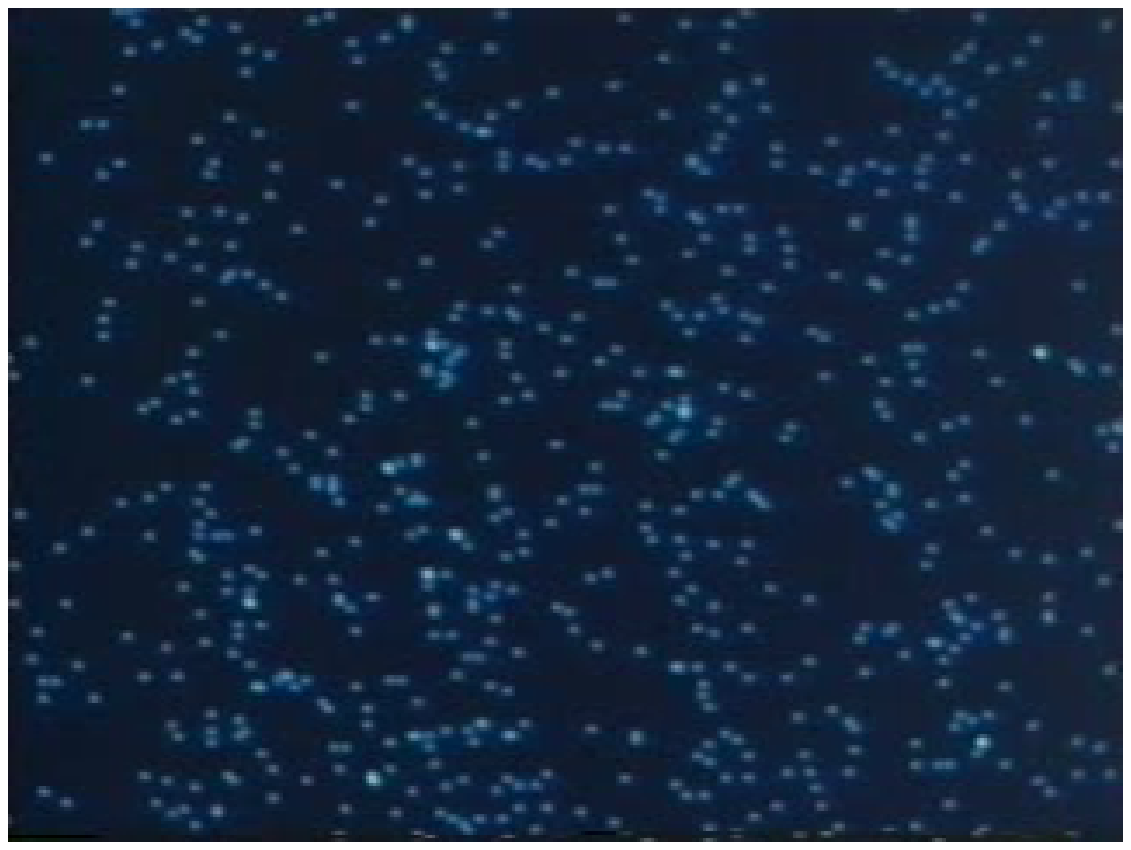}}
\end{center}

After few
hundred electrons have passed, we find a picture of random spots distributed on
the screen. However, with several thousands electrons, we get a very clear
typical interference figure.

\begin{center}
\resizebox{!}{4cm}{\includegraphics{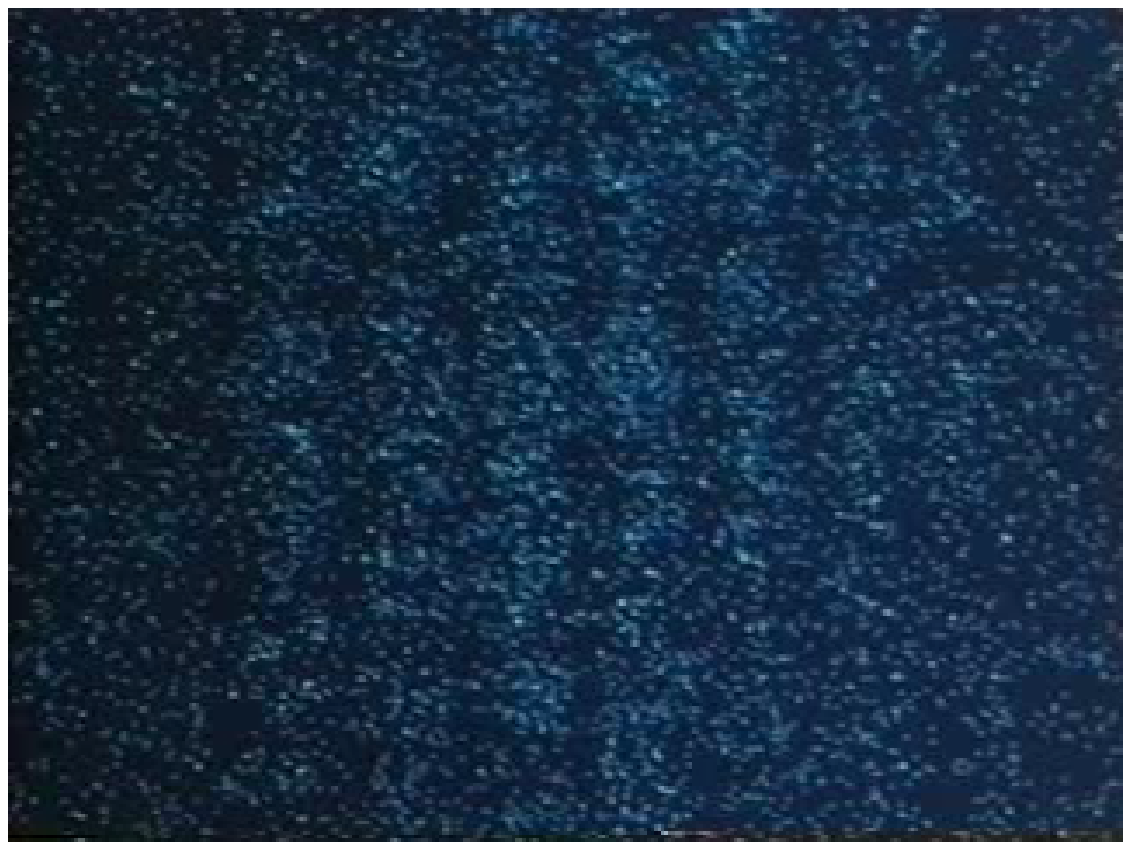}}
\resizebox{!}{4cm}{\includegraphics{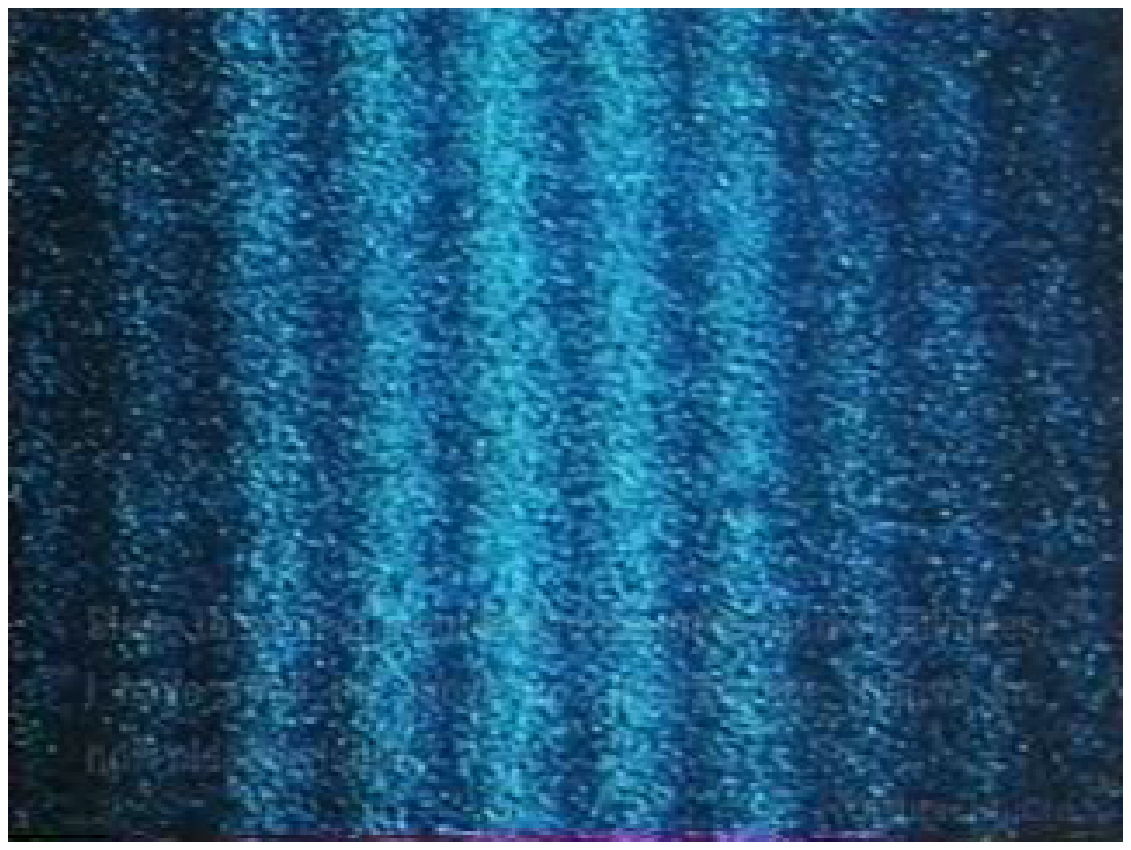}}
\end{center}

The same situation happens again if we experiment with photons (light quanta),
with an experimental arrangement that makes sure that only one photon is
present at each time.

This experiment suggests that the new theory must have a wave character  (to take
into account the interference aspects),  statistical-probabilistic
character along with an intrinsically discrete aspect. All this is quite
counter-intuitive for particles, but it is even more startling for light. Within
the classical setting we have to accept that it is not so simple to provide a
single model capable of capturing these various aspects at the same time.

From the historical point of view, things developed differently because
inconsistencies arose already in the derivation of the law for the spectral
distribution of energy density of a black-body.  Moreover, it was not possible
to account for the stability of atoms and molecules along with the detected
atomic spectra. We refer to \cite{EMS} for an account of the experimental foundations
of quantum theory  and for other background material.

The efforts of theoreticians gave rise to two alternative, but equivalent
formulations of quantum mechanics. They are usually called the Schr\"odinger
picture and the Heisenberg picture. As we are going to see in the coming
sections, the first one uses as a primary object the carrier space of states,
while the latter uses as carrier space the space of observables. 

Schr\"odinger equation has the form
\begin{equation}
  \label{eq:chrodinger}
  i\hbar \frac d{dt}\psi=H\psi
\end{equation}

The complex valued function $\psi$ is called the {\bf wave function}, it is
defined on the configuration space of the system we are considering, and it is
interpreted as a probabilistic amplitude. This interpretation requires that
$$
\int_D\psi^*\psi\  d\mu =1;
$$
i.e. because of the probabilistic interpretation $\psi^*\psi\ d\mu$ must be a probability
density and therefore $\psi$ must be required to be square-integrable. Thus wave
functions must be elements of a Hilbert space of square integrable
functions. The operator $H$, acting on wave functions, is the infinitesimal
generator of a one-parameter group of unitary transformations describing the
evolution of the system under consideration.

These are the basic ingredients appearing in the Schr\"odinger evolution
equation. The presence of the new fundamental constant $\hbar$ within the new class
of phenomena implies some fundamental aspects completely new from the previous
classical ones.  It is clear that any measurement process requires an exchange
of energy (or information) between the object we are measuring and the
measuring apparatus.  The existence of $\hbar$ requires that these exchanges cannot
be made arbitrarily small and therefore idealized to be negligible. Thus the
presence of $\hbar$ in the quantum theory means that in the measurement process we
cannot conceive of a sharp separation  between the `object' and the `apparatus' so
that we may forget of the apparatus altogether. 

We  should remark that even if the apparatus may be described classically, it is
to be considered as a quantum system with a quantum interaction with the object
to be measured. Moreover, in the measuring process, there is an inherent
ambiguity in the `cut' between what we identify as the object and what we
identify as apparatus \cite{Hei:1958,Hei:1971}. 

The problem of measurement in quantum theory is a very deep one and goes beyond
the scope of these notes. We may simply mention that within the von Neumann
formulation of Quantum Mechanics (see \cite{vonNeumann}) the measurement
problem gives rise to  the so called 
`wave-function collapse'. The state vector of the system we are considering,
when we measure some real dynamical variable $A$, is projected onto one of the
eigenspaces of $A$ with some probability that can be computed.  As the scope of
these notes is only to highlight the various mathematical structures present in
the different formulations of quantum mechanics we shall adhere to the von Neumann
projection prescription.

To avoid technicalities we shall mainly work within a finite dimensional
framework, i.e. with finite dimensional Hilbert spaces. In this setting we are
going to deal with the Schr\"odinger and Heisenberg pictures and 
we shall also provide a geometrical unifying version of the two
pictures. However, to be able to consider the quantum-classical transition in a
meaningful way, we shall consider the Weyl--Wigner formalism in infinite
dimensional Hilbert spaces.

Before closing this introduction and to better put into perspective the
Schr\"odinger and the Heisenberg pictures we are going to make a few general
considerations on the minimal mathematical structure required for the
description of a physical system.

From a general point of view, we need three ingredients:
\begin{itemize}
\item a space of states, that we denote as $\mathcal{S}$,
\item a space of observables, that we denote as $\mathcal{O}$ and
\item a real valued pairing $\mathcal{O}\times \mathcal{S}\to \R$. This pairing, which
  produces a real number out of an observable and a state, represents the
  measuring operation.
\end{itemize}

In Quantum Mechanics, we have two main pictures.
\begin{itemize}
\item in the Schr\"odinger picture, $\mathcal{S}$ is associated with a Hilbert
  space $\Hil$ and the set of dynamical variables (the observables) is a
  derived concept. Observables
 are identified with self-adjoint  bounded operators on $\Hil$.
\item in the Heisenberg picture the situation is
  complementary: the set of dynamical variables (the observables) is the
  primary concept. They are assumed to be (the real part)  of  a $\C^*$-algebra
  $\mathcal{A}$. The states, on the other hand, are a derived concept defined
  as a proper subset of the set of linear functionals on $\mathcal{A}$.
\end{itemize}

It should be stressed, however, that a physical system requires, in addition to
either one of the two primary carrier spaces, a concrete realization of it to
allow us to identify the physical variables. This last requirement is often
overlooked in the literature. We can clarify this last point with a specific
example taken from classical mechanics but that applies equally well within
Quantum Mechanics.

\begin{example}
  Let us consider the carrier space for a classical system to be a phase space
  $(\R^3-\{ \vec 0\}) \times \R^3 $ equipped with a Poisson bracket. Considering coordinates
  $(\vec \xi, \vec \eta)$ we define the Poisson structure in the
  form
$$
\{ \xi_j, \xi_k\} =0\,, \quad \{ \eta_j,\eta_k\} =\lambda \epsilon_{jkl}\frac {\xi_l}{\| \xi\|^3}\,, \quad \{\xi _j, \eta_k\}=\delta_{jk} \,. 
$$

This carrier space is appropriate to describe an electron-monopole system or a
massless particle with helicity. Indeed if we set
$$
\xi_j=x_j \text{ (position) }\,, \quad \eta_j=p_j\ \text{(momentum)}\,,
$$
the resulting Poisson brackets take the form required in the electron-monopole
system.  The brackets of the momenta are thus proportional to the magnetic
field of the monopole.

If we set $\xi_j=p_j$ and $\eta_j=x_j$, on the other hand, we endow the carrier
space with the Poisson structure required to model the dynamical behaviour of a
massless spinning  particle. The cubic term in the denominator of the bracket of
two position coordinates accounts then  for the fact that a zero-rest-mass particle
cannot be reduced to rest. And the non-vanishing of these brackets is taking
into account that massless particles cannot be localized in space. 

A very similar situation prevails in the corresponding quantum situation. For a
more detailed description of these problems, the interested reader is addressed
to \cite{BaMaSkaSte:1983}.
\end{example}

In conclusion, the description of a physical system requires not only an
abstract mathematical model (a Poisson manifold, a Hilbert space, a
$\C^*$-algebra,\ldots) but also a specific realization with an identification of
the physical variables. 

For further reading see \cite{ludwig,vanderwaerden,reichenbach,
  Peres:1995,mackey2,gudder,camilleri}. 

\section{Two formulations of Quantum Mechanics}

Our goal in this section is to present the very basic ingredients of 
Quantum Mechanics, just to establish the departure point of the analysis we will
carry on in the following sections.  We will just mention the two most familiar
formulations of Quantum Mechanics, our aim being to identify the relevant
mathematical structures required for their definition. Once we know them, they
will be studied in much more detail and from the point of view of Geometry  in  the
following sections. For more details see \cite{Dix:1977,Dirac:36,Jordan:1934,manko2,weyl,fock}.

\subsection{The Schr\"odinger formalism}
 
In this formalism the carrier space is the Hilbert space of states of the
system $\Hil$, very often the space of complex square integrable functions defined on
some spatial domain $D\subset \R^n$, identified with the configuration space.  This
is the set of pure states $\S$ of our 
system represented by the wave-functions we mentioned in the introduction.
Observables are then defined as self-adjoint operators  acting
on this Hilbert space. Thus, the set of operators $\mathcal{O}$ depends, for
its definition, on the definition of the set of states. The pairing is defined
in terms of the Hermitian structure of the Hilbert space associating a real
value to the pair $(\text{pure state},\text{observable})$ as
$$
(\psi, A)\mapsto \langle A\rangle=\langle \psi , A\psi \rangle \in \R.
$$

Dynamics is defined  on this  space by means of the Schr\"odinger equation
\begin{equation}
  \label{eq:schrodinger}
  i\hbar \frac d{dt}\psi=H\psi,  \quad \psi \in \Hil,
\end{equation}
where $H$ is the Hamiltonian operator of the system and is assumed to be Hermitian. In more
technical terms, we can
consider thus a vector field corresponding to the equations of motion
$$
\frac d{dt}\psi=\frac 1{i\hbar}H\psi, 
$$
which becomes the infinitesimal generator of a one-parameter group of unitary
transformations. As we would like to concentrate our
attention on the geometrical aspects, in this paper  we will assume, for the sake of
simplicity, that the Hilbert space is finite dimensional. 

In the particular case of a one-level system, we can introduce two real
variables $q$ and $p$ to represent $\psi$ (its real and imaginary parts
respectively $\psi=\frac 1{\sqrt{2}}(q+ip)$) and the Schr\"odinger equation takes the
  form (see \cite{Dubrovin,marmovilasi,mankobeppe}):
$$
\frac d{dt}
\begin{pmatrix}
q\\
p
\end{pmatrix}
=
\frac 1 \hbar 
\begin{pmatrix}
0 & H \\
-H & 0
\end{pmatrix}
\begin{pmatrix}
q\\
p
\end{pmatrix}.
$$
We conclude thus that the description of the dynamics in terms of real
coordinates is represented by a Hamiltonian vector field. Actually this is a
general property, also valid for infinite dimensional systems: Schr\"odinger
equations of motion will be a particular Hamiltonian dynamics on some infinite
dimensional symplectic vector space.

We can elaborate a little further on this statement. If $\Hil$ denotes a
complex Hilbert space we can decompose the Hermitian product $\langle \cdot, \cdot\rangle $ into
real and imaginary parts as follows:
$$
\langle \psi, \phi \rangle =g(\psi, \phi)+i\,\omega (\psi, \phi)\,, \quad \forall \psi, \phi \in \Hil\,,
$$
where the real and imaginary parts represent an Euclidean and a symplectic
product respectively. On the associated realification of $\Hil$ (say $\Hil_\R$)
we have a complex structure $J:\Hil_\R\to \Hil_\R$ satisfying
$J^2=-\mathbb{I}$. Thus the carrier space is endowed with a K\"ahler
structure. Vector fields associated with the Schr\"odinger equation are not only
symplectic, they are also Killing vector fields or, more specifically, they are
K\"ahlerian vector fields, i.e. they preserve the K\"ahler structure. 

\subsection{Heisenberg formalism}

In this picture observables are associated with Hermitian operators. They encode 
the measurable information of the system and the dynamics must now be defined
as a flow  on this space. States are thus defined as normalized positive functionals on
Hermitian operators.

Hermitian operators do not carry an associative algebra structure (i.e. the
product of two Hermitian operators will not be, in general,
Hermitian). However, it is possible to endow the set with one  scalar product
and two binary products:

\begin{itemize}
\item The scalar product is the restriction to the set  $\mathcal{O}$ of Hermitian operators
  of the scalar product of two complex matrices defined as
$$
\langle A, B\rangle =\Tr(A^+B).
$$
In the case of Hermitian matrices this becomes
\begin{equation}
  \label{eq:scalarproduct}
  \langle A, B\rangle =\Tr(AB) \quad \forall A, B\in \mathcal{O}.
\end{equation}
\item The first binary operation is the Abelian real Jordan algebra product
$$
A\circ B:= \frac 12[A,B]_+=\frac 12 (AB+BA)=\frac 14 \left ((A+B)^2-(A-B)^2 \right )\,.
$$
Let us recall, for completeness, the definition of Jordan algebra:
\begin{definition}
  A (non-associative) algebra $(\mathcal{A}, \cdot)$ is called a {\bf Jordan algebra} if
  the composition law  is commutative and for any two arbitrary elements
  $A,B\in \mathcal{A}$, $    (AB)A^2=A(BA^2)$.
\end{definition}

With this definition we can conclude
\begin{lemma}
 $(\O, \circ)$ is  a Jordan algebra. 
\end{lemma}
\proof{
The commutativity is obvious. The second condition 
 follows from the associativity of the original product:
$$
[[A,B]_+,A^2]_+= (AB+BA)A^2+A^2(AB+BA)=(ABA^2+BA^3+A^3B+A^2BA)
$$
$$
[A,[B,A^2]_+]_+=A(BA^2+A^2B)+(BA^2+A^2B)A=(ABA^2+A^3B+A^3B+A^2BA)
$$
}

The map $\{ A,B,C\}=(A\circ B)\circ C-A\circ (B\circ C)$ is called the {\bf associator} of the
algebra, and the algebra is associative if and only if the associator is
identically zero.

\item The second binary structure is a Lie algebra structure
$$
[A,B]_-=\frac 1{i\hbar}(AB-BA),
$$
which comes from the fact that for any  Hermitian operator $A$, $-iA$ is an infinitesimal
generator of the unitary group.

Therefore, multiplying each element in the set by the imaginary unit we get the
Lie algebra of the unitary group.

\end{itemize}

\begin{proposition}
The scalar product \eqref{eq:scalarproduct} is also invariant with respect to this new
product, and we have:
\begin{equation}
  \label{eq:compa}
  \langle [A,B]_{-}, C\rangle =\langle A, [B,C]_{-}\rangle \,, \qquad  \langle [A,B]_{+}, C\rangle =\langle A, [B,C]_{+}\rangle\,.
\end{equation}

Moreover we also have the compatibility relation
\begin{equation}
  \label{eq:compatb}
  [A, B\circ C]_{-}=[A,B]_{-}\circ C+B\circ [A, C]_{-}\,
\end{equation}  
i.e. $\ad_A$ is a derivation of the Jordan algebra for any $A\in \mathcal{O}$.
\end{proposition}
\proof{
These properties follow directly from the definitions.   
}

Actually these two structures can be combined together in the notion of
Lie--Jordan algebra (see \cite{Emch} for details).

Now we can proceed to define dynamics on this algebra. It is
introduced by means of the {\bf Heisenberg equation}, which makes use of the
skew-symmetric structure of the algebra:
\begin{equation}
  \label{eq:heisenberg}
   \frac d{dt}A=[A,H]_- \quad A\in \mathcal{A}
\end{equation}
where $H$ is the Hamiltonian of the system.

\begin{remark}
  The equations of motion written in this form are necessarily derivations of
  the two products (i.e. a derivation of the Lie--Jordan product) and can be
  considered hence `intrinsically Hamiltonian'. In the Schr\"odinger picture,
  if the vector field is not anti-Hermitian, the equation still makes sense,
  but the dynamics is not K\"ahlerian.
\end{remark}

\section{Geometric Quantum Mechanics I: The Sch\"odinger picture}
Our purpose in this section is to present Quantum
Mechanics from a geometric perspective. We choose to do it in the case of finite
dimensional Hilbert spaces (i.e. systems with a finite number of energy
levels, for instance) because of their simplicity, although all the objects
that we are going to introduce make sense in general for  infinite dimensional
Hilbert spaces as well. Further details can be found in
\cite{ashschi:1999,cantoni:1975,cantoni:1977,Mack:1962,cirelli3,cirelli2,bloch,
  heslot,rowe,field,brody,strocchi,BCG:1991,cantoni3,cantoni4,cantoni5,
  mackeylibro,darius,anandan,cirelli4,benvegnu}.

\subsection{The Hilbert space as a real differentiable manifold}

Thinking in terms of the Schr\"odinger picture, we know that the set of pure states
$\S$ is associated with  a Hilbert space. Let us study in detail the geometrical
objects which play a role in the definition of the dynamics within the Schr\"odinger
picture. 

We want to consider the space $\S$ as a differentiable manifold instead of a
linear space. We can consider the complex 
vector space as a real vector space (i.e. a `realification')  if we consider 
the natural complex structure $J$ ($J^2=-1)$ of the Hilbert space. 
But to consider a differentiable structure implies
that we have to associate tensorial objects with the 
vectors and linear maps which we have studied so far.

\subsubsection{The tensors}

\begin{itemize}
\item The first task is the association of vectors of $\Hil$ with vector fields on
  the manifold.  Being a linear space, $\Hil$ can be identified with the
  tangent space at any point, and hence we can write $T\Hil\sim \Hil\times \Hil$. Thus
  it makes sense to consider, for an element 
  $\eta\in \Hil$ the vector field $X_\eta$ defined as a section of $T\Hil$:
$$
 X_\eta :\psi \mapsto  (\psi, \eta)\,.
$$ 

This vector field acts on a function $f$ as:
$$
X_\eta(f)(\psi)=\frac{d}{dt}f(\psi+t\eta)|_{t=0}.
$$
Besides, these constant sections define a separating set in the Hilbert space.
\item Let us consider the K\"ahler structure on the Hilbert space. Let 
$\langle \psi_1, \psi_2\rangle\in \C$ denote  the scalar product
of two vectors $\psi_1$ and $\psi_2$, 
 and  consider the structure of real manifold. The scalar product above is written as 
$
\langle \psi_1, \psi_2\rangle=g(X_{\psi_1}, X_{\psi_2})+i\,\omega (X_{\psi_1}, X_{\psi_2}),
$
where $g$ is now a symmetric tensor and $\omega $ a skew-symmetric one. The properties
of the Hermitian product ensure that:
\begin{itemize}
\item the symmetric tensor is positive definite and non-degenerate, and hence
  defines a Riemannian structure on the real vector space.
\item the skew-symmetric tensor is also non degenerate, and is closed with
  respect to the natural differential structure of the vector space. Hence, the
  tensor is a symplectic form.
\end{itemize}

As the inner product is sesquilinear, it satisfies
$$
\langle \psi_1, i\psi_2\rangle =i\langle \psi_1, \psi_2\rangle, \qquad \langle i\psi_{1}, \psi_{2}\rangle =-i\langle
\psi_{1},\psi_{2}\rangle. 
$$
This  implies
$$
g(X_{\psi_1}, X_{\psi_2})=\omega (JX_{\psi_1}, X_{\psi_2}),
$$
or, equivalently, that the triple $(J, g, \omega )$ defines a K\"ahler structure.

These two tensors $g$ and $\omega$ are in a covariant form . We can also define their
contravariant forms by considering the dual vector space $\Hil^*$ identified
wth $\Hil_\R$, for instance
via the metric $g$ (which is non-degenerate). The association of vectors of
$\Hil$ with vector fields can be extended to associate also 1-forms with the
elements of $\Hil^*$. We will have then an assignment $\Hil^*\ni \widetilde \psi\mapsto
\alpha_{\widetilde \psi}:\phi \mapsto (\phi, \widetilde \psi)$, i.e. we write $T^*\Hil\sim \Hil\times
\Hil^*$. In this way we define the contravariant tensors $G$ and $\Omega$, which
allow us to define a scalar product on $\Hil^*$ as:
$$
\langle \widetilde \psi_1, \widetilde \psi_2 \rangle=G(\alpha_{\widetilde \psi_1},
\alpha_{\widetilde \psi_2})+i\Omega (\alpha_{\widetilde \psi_1},
\alpha_{\widetilde \psi_2})\quad \forall \widetilde \psi_1, \widetilde \psi_2  \in \Hil^*\,.
$$

If we select an orthonormal basis $\{e_1,\ldots,e_n\}$ for $\mathbb{C}^n$, 
we may define coordinates by setting 
$\braket{e_k}{\psi}=z_k(\psi)=\frac 12(q_k+i\, p_k)(\psi)$, and we have used Dirac's
 notation for bras and kets.

In these
coordinates we have a  contra-variant version of the Euclidean structure given by 
$
G=\sum_{k=1}^n\left(\pd{}{q_k}\otimes \pd{}{q_k}+\pd{}{p_k}\otimes \pd{}{p_k}\right)
$
and the Poisson tensor
$
\Omega=\sum_{k=1}^n\left(\pd{}{q_k}\land \pd{}{p_k}\right)
$
while the complex structure has the form 
$J=\sum_{k=1}^n\left(\pd{}{p_k}\otimes d{q_k}+\pd{}{q_k}\otimes d{p_k}\right)
$.

In terms of complex coordinates the Hermitian structure has the form 
$h=\sum_{k=1}^n d\bar z_k\otimes  dz_k$. The corresponding contra-variant form is given by 
$$G+i\,\Omega=\sum_{k=1}^n\left(\pd{}{q_k}-i\pd{}{q_k}\right)\otimes
\left(\pd{}{q_k}+i\pd{}{q_k}\right)=4\,  \sum_{k=1}^n \pd{}{z_k}\otimes
\pd{}{\bar z_k}\,.
$$

We may now define binary products on functions by setting 
$$
\begin{array}{rcl} \{f_1,f_2\}&=&{\displaystyle\sum_{k=1}^n\left(\pd
  {f_1}{q_k}\,\pd{f_2}{p_k}-
\pd {f_1}{p_k}\,\pd{f_2}{q_k}\right)}\,,\cr
\{f_1,f_2\}_+&=&{\displaystyle\sum_{k=1}^n\left(\pd
  {f_1}{q_k}\,\pd{f_2}{q_k}+
\pd {f_1}{p_k}\,\pd{f_2}{p_k}\right)}\,,\cr
\braket{f_1}{f_2}&=&4\ {\displaystyle\sum_{k=1}^n\pd{f_1}{z_k}\,\pd{f_2}{\bar z_k}}\,.
\end{array}
$$ 
\end{itemize}

\subsubsection{Additional  tensor fields}
In addition, we can consider the linear structure of the Hilbert space and
associate with it  the Liouville vector field:
\begin{equation}
  \label{eq:liouville} 
 \Delta:\Hil\to T\Hil\,, \quad \psi\mapsto (\psi, \psi),
\end{equation}
which, as usual, allows us to define homogeneous polynomial functions: a
function $f\in C^\infty (\Hil)$ is homogeneous of degree $k$ if $\Delta(f)=kf$. Combining
this tensor with the complex structure, it is possible to define a new vector field
\begin{equation}
  \label{eq:gamma}
  \Gamma=J(\Delta).
\end{equation}

$\Gamma$ and $\Delta$ commute and therefore generate an integrable distribution which
defines a foliation on the Hilbert space.

These two tensors also help us to define a way of associating a tensor to any
operator acting on $\Hil$. There are several ways to do it, some are more
immediate than others:
\begin{itemize}
\item We can associate a quadratic function $f_t$ to any constant symmetric 2-tensor field
  $t$ in  the form:
  \begin{equation}
    \label{eq:funcm}
    f_t(\psi )=\frac 12 t(\Delta,\Delta )(\psi)=\frac 12 t(\psi, \psi);
  \end{equation}
and similarly for higher order tensors.
\item Skew-symmetric 2-tensors $\gamma$ are transformed into functions in a similar
  way, using also the complex structure:
  \begin{equation}
    \label{eq:skew}
    f_\gamma(\psi )=\frac 12\gamma(\Delta, \Gamma )(\psi)\,.
  \end{equation}
When we consider as skew-symmetric tensor the symplectic form $\omega $, the
resulting function is the Hamiltonian function generating the one-parameter
group of unitary transformations which defines the multiplication by a phase. 

\item Any linear operator $A:\Hil\to \Hil$ can be identified with:
  \begin{itemize}
  \item a (1:1) tensor field
    \begin{equation}
      \label{eq:Ta}
      T_A:T\Hil\to T \Hil \qquad T_A:(\phi, \psi)\mapsto (\phi, A\psi)\,,
    \end{equation}
    \item or two different  vector fields
      \begin{equation}
        \label{eq:Xa}
        X_A=T_A(\Delta):\Hil \to T\Hil \qquad X_A:\psi\mapsto (\psi, A\psi)\,,
      \end{equation}
      and 
      \begin{equation}
        \label{eq:Ya}
        Y_A=T_A(J(\Delta)):\Hil\to T\Hil,       \qquad Y_A:\psi \mapsto (\psi, JA\psi).
      \end{equation}
     If $A$ is Hermitian, then $X_A$ corresponds to a gradient vector field
with respect to the K\"ahler structure, while $Y_A$ corresponds to the
Hamiltonian vector field associated to the evaluation function of the
operator (i.e. $\psi\mapsto \langle \psi, A\psi\rangle $). These associations have
 different properties:
\begin{itemize}
\item the mapping $A\mapsto T_A$ is an isomorphism of associative
  algebras (and as a result also with respect to the Lie algebra structure).
As we are interested in the `realification' of operators acting on the
complex vector space, we shall restrict our considerations to
 tensors satisfying $T_AJ=JT_A$. 
\item the mappings $A\mapsto X_A$ or $A\mapsto Y_{A}$ on the other hand are only isomorphisms of Lie
  structures, and the properties of the associative product of operators is lost.
\end{itemize}
\end{itemize}
\item  Occasionally, to make easier the comparison with the usual formalism, we
  consider the space $\S$ as a real manifold but, at each point $\psi$, we may consider
  $T_\psi\S$  as a complex vector space.  In this case, vector fields would have a
  real and an imaginary part. However, even when this notation might be
  misleading, we shall always be considering the derivations on $\S$ in the real
  sense. Hence, we shall not be considering derivations with respect to complex
  variables and, as  a result, we do not need to consider complex analiticity
  for our functions. We will have, though, complex valued functions arising as
  the contraction of complex valued vector field with complex values one
  forms.

By using the `mixed' point of view,  it is also possible to associate a complex
  valued quadratic function on $\Hil$ to any linear transformation $A:\Hil\to
  \Hil$ by defining
  \begin{equation}
    \label{eq:f_A}
    2f_A(\psi)=g(\Delta, T_A(\Delta))(\psi)+i\,\omega (\Delta, T_A(\Delta))(\psi)=\langle \psi, A\psi \rangle \,.
  \end{equation}

Given the quadratic  function $F\in \F(\Hil )$,
associated to the operator $\hat F$, the 1-form $dF$ 
acts on a vector field $X_\eta$ as 
$$
dF(X_\eta)(\psi)=X_\eta(F)(\psi)=\frac 12 \frac{d}{dt}\langle \psi + t\eta, \hat F(\psi +t\eta )\rangle \mid _{t=0}\,.
$$

The Hamiltonian
vector field corresponding to $F$ by $\omega$:
\begin{eqnarray}
dF(X_\eta)(\psi)&=&\frac 12 \langle \psi, \hat F\eta\rangle +\frac 12 \langle \eta , \hat F\psi\rangle =g(X_{\hat F}, X_\eta )(\psi)\cr&=&
\omega (Y_F, X_\eta )(\psi)=(i_{Y_F}\omega  )(X_\eta )(\psi)\,,\nonumber
\end{eqnarray}
where we used the relation between the Riemannian and the symplectic K\"ahler
forms and the definition of the vector fields.

In conclusion tensor fields on the real manifold $\S$  will be  considered as
modules over complex-valued functions on $\S$.
\end{itemize}

Please notice that the above definitions are intrinsic and can be applied
whenever the tensors used are available. As a result, it is also possible to
define these objects at the level of infinite dimensional Hilbert spaces.


\subsubsection{Observables as quadratic functions}
Our geometrization procedure has allowed to replace operators with complex
valued functions $A\mapsto f_A$. This association is clearly injective, but it is not
onto, i.e. there are functions on $\Hil$ which are not quadratic. The
association is obviously linear but the image is not closed under the pointwise
product (the product of two quadratic functions is not quadratic but
quartic). Therefore the product  cannot be the image of an operator. Thus the
pointwise product does not allow to transfer the associative product of
operators to the set of quadratic functions. However, we might consider a
non-local product, inner in the space of quadratic functions, and defined as
\begin{equation}
  \label{eq:nonlocal}
(  f_A\star f_B)(\psi)=(f_{AB})(\psi)\,.
\end{equation}

This product is not commutative and non-local. It requires, however, that we
start with operators, or their tensorial versions as  $(1,1)$ tensor fields. 
It seems advisable to describe this
product only in terms of the tensors already available on the real differentiable
manifold $\S$.

We notice first that if $A$ and $B$ are Hermitian, with associated real valued
quadratic functions $f_A$ and $f_B$, the product $f_A\star f_B$ need not be a real
valued function (for the product of two Hermitian operators is not Hermitian,
in general).  Using the fact that these functions are quadratic, we may
consider the quantity $G(df_A, df_B)$, where $G$ is the contravariant form of
the metric tensor. This combination is clearly a quadratic function, because we
know that $G$ satisfies $L_\Delta G=-2G$. By straightforward computations, we can
obtain that 
$$
G(df_A, df_B)=f_{AB+BA}\,.
$$

In a similar way, by using the skew-symmetric tensor $\Omega$, we obtain
$$
\Omega(df_A, df_B)=-if_{[A,B]}=f_{[A,B]_-}\,.
$$

Thus the Lie--Jordan algebra structure on the space of real quadratic functions
can be extracted from the Hermitian tensor. For later use (when we consider the
complex projective space) it is convenient to characterize these functions
defining the Lie--Jordan structure without using the notion of quadratic
function (for it does not make sense on nonlinear spaces). It is possible to show
that:
\begin{lemma}
  Given a function $f$, the Hamiltonian vector field $X_f=\Omega(df, \cdot )$
  preserves the metric tensor $G$, i.e. $L_{X_f}G=0$ if and only if $f$ is a
  quadratic function associated with some Hermitian operator.
\end{lemma}

Thus, a subset of functions in $\F(\Hil)$ can be selected and defines a
Lie--Jordan algebra with the 
tensors $G$ and $\Omega$ if and only if they are real quadratic.  Moreover, if we
consider this subset of functions we get
$$
G(df_A,df_B)+i\Omega(df_A, df_B)=f_A\star f_B.
$$
By linearity, we can extend these operations to complex valued functions.
Then, this operation defines a $\C^*$-algebra structure on the space of
complex valued functions whose real and imaginary parts are associated with
Hermitian operators. The norm of this $\C^*$--algebra is given by the usual
$\mathrm{sup}$ norm, i.e. the supremum of the values that the operator takes on
normalized states.

In this way, our geometrization procedure has reproduced the algebra of
observables in terms of real valued functions on $\S$.

\subsubsection{Transformations}
As we have already remarked, the evolution of a quantum system defines
a one-parameter group of  transformation of $\S$ which preserve the 
K\"ahler structure. All transformations preserving the K\"ahler structure  form the
set of unitary transformations of 
$\Hil$, and in the case of finite dimensional Hilbert spaces ($\dim_\C \Hil=N$),
they provide a realization of the unitary group $U(N)$.

In order to represent these operators in our setting, we can consider the set
of Hermitian operators and use the vector field association
\eqref{eq:Xa}. 
The one parameter group of unitary transformations associated with the
Hermitian operator $A$ is
\begin{equation}
  \label{eq:tra}
  U(\alpha)=e^{-i\alpha A/\hbar}.
\end{equation}

\subsection{The complex projective space}

The probabilistic interpretation of states assigns a physical meaning only to
`normalized wave-functions by means of the probability densities $\psi^*\psi 
d\mu$'. Thus the meaningful physical space is the complex projective space
associated to the Hilbert space $\Hil$. Then, the
next step is to induce the geometrical structures we have considered above onto the
complex projective space.

The usual way to define the complex projective space is by means of equivalence
classes. The complex vectors $\psi_1$ and $\psi_2$ are considered to be
equivalent if there exists a nonzero complex number $\lambda$ such that
$\psi_2=\lambda\psi_1$. We can denote the equivalence class by one of its
representatives, say $[\psi]$. If we write $\lambda=\rho e^{i\alpha}$ ($0<\rho\in \R ,\alpha  \in \R$) we see immediately
that equivalence classes are orbits of the group $S^1\times \R_+$. Thus, removing
the zero vector, the complex projective space is the set of orbits of that
group acting on  $\Hil-\{ 0\} $. The infinitesimal generators of the action
can be easily determined: $\Delta$ is the generator of the modulus part
(dilations) while $\Gamma=J(\Delta)$ is the generator of the phase change. If we consider
the vector space endowed with its canonical K\"ahler structure $(J, g, \omega)$, we
can further characterize $\Gamma$ as being the Hamiltonian vector field
corresponding to the function $\frac 12 g(\Delta, \Delta)$:
\begin{lemma}
  $\Gamma$ is the Hamiltonian vector field corresponding to the quadratic function
  associated with the identity operator.
\end{lemma}
\proof{
Having considered the function $\frac 12 g(\Delta, \Delta)$, we have
$$
d\left (\frac 12 g(\Delta, \Delta)\right )(X_\eta)=g(\Delta, X_\eta )=\omega(J\Delta, X_\eta )=i_{\Gamma }\omega(X_\eta) \,.
$$
}

The vector fields $\Delta$ and $\Gamma$ commute and therefore define an involutive
distribution, hence we can consider the foliation defined by such
vector fields. The corresponding quotient manifold,
is again the complex projective space
$\mathcal{PH}$.

This remark allows us to consider projectable tensorial quantities. For
instance, we have:
\begin{lemma}
For each Hermitian operator $A$,   the expectation value function, defined as
  \begin{equation}
    \label{eq:expval}
    e_A(x)=\frac{\langle x, Ax \rangle}{\langle x, x\rangle}\,,
  \end{equation}
is invariant under $\Delta$ and $\Gamma$.
\end{lemma}
\proof{
  The invariance under the Liouville vector field is immediate. The invariance
  under $\Gamma$ follows from the fact that
$$
\Gamma(e_A)=\{ e_{\mathbb{I}}, e_A\}=e_{[\mathbb{I},A]}=0\,.
$$
}

But hence we proved:
\begin{theorem}
  The space  of  the expectation
  values functions of Hermitian operators projects onto the quotient space defined by the
  foliation generated by $\Delta$ and $\Gamma$.
\end{theorem}

A general operator, dynamical variable,  can be decomposed into real and
imaginary parts, both parts given by Hermitian operators. Therefore we can say
that both parts project onto the complex projective space.

\subsubsection{Eigenvalues and eigenstates}
In the `geometrized' Hilbert space description we have to recover now the
description of the `eigenvectors' and `eigenvalues'. 

To this aim is appropriate to introduce expectation values associated with
Hermitian operators, $A=A^+$: 
$$
e_A(\psi)=\frac{\langle{\psi},{A\psi}\rangle}{\langle{\psi},{\psi}\rangle}\,.
$$

We find that:

\begin{enumerate}
\item Critical points of $de_A$  correspond to the eigenvectors of $A$.
\item Values of $e_A$ at critical points are the corresponding eigenvalues of $A$.
\end{enumerate}

\begin{remark}
   Critical points of $de_A$ coincide with the critical
 points of the corresponding Hamiltonian vector field $\Omega(de_A)$ or 
of the corresponding
gradient vector field $G(de_A)$.
\end{remark}

\subsubsection{Observables on the complex projective space}
We have seen that in our geometrization the algebra of observables can be
recovered in terms of functions with the help of the contravariant tensors $G$
and $\Omega$. We noticed that expectation value functions are projectable and
therefore it makes sense to consider the binary products $G(de_A, de_B)$ and
$\Omega(de_A, de_B)$. Unfortunately, because of $L_\Delta G=-2G$ and $L_\Delta \Omega=-2\Omega$ the
result of those operations on projectable functions will not be
projectable. Thus, in order to make them 
inner operations, we may use a conformal factor for both tensors, as, for
instance $G_P=\langle \psi, \psi\rangle G$ and $\Omega_P=\langle \psi, \psi \rangle \Omega$. In this way we would
define projectable tensors, and hence inner products of projectable functions. 
A new problem arises though: if we define the bracket corresponding to the
skew-symmetric part
$$
\Omega_P(df, dh)(\psi)=\{ f, h\} _P(\psi )=\langle \psi, \psi\rangle \{ f, h\} (\psi),
$$
this new bracket does not satisfy the Jacobi identity. Indeed, to make it to
satisfy the Jacobi identity we have to consider a Jacobi bracket instead of a
Poisson one, in the form
\begin{equation}
  \label{eq:jacobibracket}
  [f,h]=\{ f,h\}_P+fL_{X}h-hL_Xf,
\end{equation}
where $X=\Omega(d\langle \psi, \psi \rangle)$ is the Hamiltonian vector field associated with the
function $\langle \psi, \psi\rangle $. Now, for the expectation values function we find that
$$
[e_A, e_B]=\{ e_A, e_B\} _P\,,\qquad \forall A,B,
$$
because the function  $\langle \psi, \psi\rangle $ is a central element for the subalgebra
generated by the expectation value functions. Therefore, the use of the
conformal tensors $G_P$ and $\Omega_P$ allows us to define a $\C^*$-algebra
structure on the space of expectation value functions, which are projectable
onto the complex projective space. 

By explicit computation it is possible to show that
$$
e_A\star e_B=e_{AB}=G_P(de_A, de_B)+i\Omega_P(de_A, de_B)+e_Ae_B\,.
$$

If we consider the projection $\pi:\Hil_0=\Hil-\{ 0\} \to \mathcal{P}\Hil$ we may identify
$\pi^*(\F(\mathcal{P}\Hil))$ with the subalgebra of $\F(\Hil)$ satisfying the
conditions 
$$
df(\Delta)=0; \quad df(J(\Delta))=0.
$$

Within this subalgebra we may further restrict to those functions $f$ such that
$Y_f=\Omega(df)$ satisfy $L_{Y_f}G=0$. This subset of functions gives rise to a
$\C^*$-algebra. 

\begin{remark}
  The symmetric product of the expectation value function associated with a
  given Hermitian operator is
$$
G_P(de_A, de_A)=\frac{\langle \psi , A^2\psi\rangle}{\langle \psi,
  \psi\rangle}-\frac{\langle \psi , A\psi\rangle\langle \psi ,
  A\psi\rangle}{\langle \psi, \psi\rangle^2}. 
$$
It represents the dispersion of the main value of the observable corresponding
to the operator $A$ when measured in the pure state $\psi$. Therefore the square
of the corresponding Hamiltonian vector field is strictly related to the
uncertainty in the measurement of $A$ in the pure state $\psi $.
\end{remark}

\subsection{Quantum dynamics}

Let us consider now the dynamics. On the Hilbert space considered as a real
differential manifold it is possible to rewrite Schr\"odinger equation by using 
the tensors introduced above so as to become
$$
\dot \psi=-J\hat H\psi \,,
$$
where we took $\hbar=1$, and $J$ is the complex structure.

The solutions of this equation corresponds to the flow of the vector field
$$
Y_H=T_H(J(\Delta)).
$$

Having introduced a vector field to describe the dynamics we may now state a
few properties:
\begin{itemize}
\item[i)] $Y_H$ is Hamiltonian with Hamiltonian function $f_H=\frac 12 \langle \psi, H\psi
  \rangle $. It is projectable onto the complex projective space.
\item[ii)] The vector field associated with the Hamiltonian function $e_H=\frac {\langle \psi, H\psi
  \rangle} {\langle \psi, \psi\rangle}$ projects onto the complex projective space. It projects onto
the same vector field associated with $Y_H$.
\item[iii)] Critical points of $e_H$ (projected) on $P\Hil$ correspond to the eigenvectors
  of $H$ and the values of $e_H$ at those points correspond to the eigenvalues.
\item[iv)] The Lie algebra of symmetries for the dynamics is generated by the
  expectation-value-functions   associated to Hermitian operators commuting
  with $\hat H$. 
\item[v)] The dynamics can be written in terms of Poisson bracket and defines a
  derivation for the $\star$-product.
\end{itemize}

\section{Geometric Quantum Mechanics II: the Heisenberg picture} 

\subsection{Introduction}
Following the algebraic approach advocated by Segal \cite{Segal:1947} and
Haag and Kastler \cite{HaagKast:1964}, we consider the space of observables as
the collection of all the self-adjoint elements of a $\C^{*}$-algebra with
identity element \cite{Thir:1981}. 

States are identified as the elements of the convex body 
$
\mathcal{S}=\{\phi\in\mathcal{A}^*\mid \phi(A^*A)\geq 0\,, \forall A\in
 \mathcal{A}; \phi(\uno)=1\}
$. A state is pure if it cannot be written as a convex combination of other
two states. 

It is not difficult to show in the finite-dimensional case, for the
infinite-dimensional case it is a theorem by Gleason \cite{gleason}, that any
state can be written in the form  
$
\phi  (A)={\rm Tr\,}\rho_\phi A\,.
$
When $\mathrm{dim} \Hil=\infty$ density states are charactaerized by the property
of being normal states on the von Neumann algebra of bounded operators
(i.e. completely additive states).

Moreover, $\rho_\phi$ is a no-negatively defined operator of ${\mathfrak{gl}}(\Hil)$,
i.e. those  $\rho_\phi\in {\mathfrak{gl}}(\Hil)$ which can be written in the form 
$\rho_\phi=T^+T$ for some $T\in {\mathfrak{gl}}(\Hil)$ and, in addition, satisfy 
${\rm Tr\,}\rho_\phi=1$. Thus, $\mathcal{S}$ is a convex body in the
 affine hyperplane in ${\mathfrak{u}}^*(\Hil)$, determined by the equation
 ${\rm Tr\,}\rho_\phi=1$.
The tangent space to this 
affine hyperplane at a point is therefore identified with
the space of traceless Hermitian operators, and it is in a one-to-one correspondence with the
 Lie algebra of the group $SU(\Hil)$.

\subsection{The geometrical description of the Heisenberg picture}

Now, we are
going to see how it is possible to obtain the Heisenberg description in
geometrical terms.

A specific way to `geometrize' the Lie algebra structure of $\u(\Hil)$ is to
associate with it a linear Poisson tensor on the dual vector space
$\u^*(\Hil)$ as follows. As
$\Hil$ is assumed to be finite-dimensional, we can
identify $\u(\Hil)$ with the space of real valued linear functions on its dual
space, i.e. $\u(\Hil)= \mathrm{Lin}(\u^*(\Hil), \R)$, and we set, for any pair of
linear functions on $\u^*(\Hil)$ defined by the  two elements $u,v\in \u(\Hil)$:
\begin{equation}
  \label{eq:Lie}
  \{ \hat u,\hat v\}=\widehat{[u,v]}\,,
\end{equation}
where the commutator on the right hand side is computed by thinking of $u,v$ as
elements of the Lie algebra  $\u(\Hil)$, and the left hand side is to be read
as a linear function on $\u^*(\Hil)$. We will use the `hat' to denote the
elements of $\u(\Hil)$ seen as linear functions on the dual $\u^*(\Hil)$.
 We are implicitly using here the property that
the vector space $\u(\Hil)$ is isomorphic to its bi-dual, which holds for vector
spaces which are reflexive, in particular finite dimensional ones. Then, we
have:
\begin{proposition}
  Let $\O$ be the space of observables of a finite level quantum system. Then,
  $\O^*$  can be endowed with a Poisson structure.
\end{proposition}

Having replaced the Lie algebra structure with the Poisson tensor associated
with the Poisson bracket on $\u^*(\Hil)$, we are now able to perform also
nonlinear transformations on the Poisson manifold. In
this sense we speak of the `geometrization' of the algebra structure of the
vector space $\u(\Hil)$. 

If we denote by $\hat A$ and $\hat B$ the linear functions on $\u^*(\Hil)$
corresponding to elements $A,B\in \u(\Hil)$, we can define the Poisson bi-vector $\Lambda$
as:
\begin{equation}
  \label{eq:poissontensor}
  \Lambda(d\hat A, d\hat B)(\xi)=\{ \hat A, \hat B\} (\xi)=\xi([A,B])=\frac i2 \xi
 (AB-BA)\,, \quad \xi \in \u^*(\Hil)\, ,
\end{equation}
where we used the scalar product of the Lie algebra.

Hence we recover the well-known Kirillov--Konstant--Souriau Poisson
tensor on the dual of any Lie algebra.

By using a similar procedure we may also `geometrize' the Jordan algebra
structure on the space of observables. Again we set:
\begin{equation}
  \label{eq:jordantensor}
  \mathcal{R}(\xi)(d\hat A, d\hat B)=\xi([A,B]_+)=
\frac i2 \Tr \xi (AB+BA)\,, \quad \xi \in \u^*(\Hil)\, ,
\end{equation}
where use is made of the relation $\xi (A)=\frac i2 \Tr (\xi A)$.

These two tensor fields can be put together to form a complex tensor field:
\begin{equation}
  \label{eq:assocuativetensor}
  (\mathcal{R}+i\Lambda)(\xi)(d\hat A, d\hat B)=(\widehat{AB})(\xi )=\xi (AB) =
\Tr(\xi AB)\,, \quad \xi \in \u^*(\Hil)\,.
\end{equation}
By using this tensor field we can define a $*$-product in the form
$$
(\hat A* \hat B)(\xi)=\xi (AB)=(\mathcal{R}+i\Lambda)(\xi)(d\hat A, d\hat B).
$$

In this context, the compatibility condition of the Lie and the Jordan
structures can be simply stated by saying

\begin{proposition}
The Hamiltonian vector fields
associated with observables (i.e. real linear functions on $\u^*(\Hil)$) are
infinitesimal symmetries for the tensor field $\mathcal{R}$ associated with the
Jordan structure, and therefore derivations for the $*$-product.  
\end{proposition}
\proof{
  It is a direct consequence of the compatibility between both brackets,
  summarized in (\ref{eq:compatb}).
}

\begin{remark}
Our `geometrization' carries along the possibility of performing nonlinear
transformations because we have replaced the algebraic structures on the linear
space $\u(\Hil)$ with tensorial objects on the manifold $u^ *(\Hil)$. It should
be remarked, however, that now
we have the possibility of two different associative products on linear
functions  on $\u^*(\Hil)$:
\begin{itemize}
\item the point-wise product $(\hat A\cdot \hat B)(\xi)=\hat A(\xi)\hat B(\xi)$, which
  gives a quadratic function out of two linear ones and
\item a non-local product $(\hat A\star \hat B)(\xi)=\widehat{AB}(\xi)$. In this case we
  obtain a linear function as the product of other two linear ones, but in general it
  will be a complex valued function even if the factors were real ones. This
  result has to do with the fact that the product of two Hermitian operators is
  not Hermitian and therefore it gives rise to real and imaginary parts.
\end{itemize}
\end{remark}

\begin{example}{\cite{Grabowski:2005}}
  At this point it may be adequate to give a simple example of the objects
  introduced so far. Let us consider the Lie algebra $\mathfrak{u}(2)$ of $2\times 2$ Hermitian
  matrices corresponding to a spin $1/2$ physical system. We introduce an
  orthonormal basis with respect to the scalar product on the algebra. We
  set thus:
  \begin{equation*}
    U=\begin{pmatrix}
1& 0\\
0&1
\end{pmatrix},
\quad  
    X=\begin{pmatrix}
0& 1\\
1&0
\end{pmatrix},
\quad 
    Y=\begin{pmatrix}
0&- i\\
i&0
\end{pmatrix},
\quad
    Z=\begin{pmatrix}
1& 0\\
0&-1
\end{pmatrix},
\quad \,,
  \end{equation*}
and also the associated linear functions
$$
\hat X=x,\quad \hat Y=y, \quad \hat Z=z, \quad \hat U=u 
$$
where the functions are to be understood as $z(A)=\frac i2 \Tr(ZA)$, and so on,
for any $A\in {\goth{u}}(2)$. In these coordinates, the Poisson tensor field is given by
\begin{equation*}
  \Lambda=2\left (x\frac\partial{\partial y}\land  \frac \partial{\partial z}+y\frac\partial{\partial z}\land \frac \partial{\partial
      x}+z\frac\partial{\partial x}\land \frac \partial{\partial y}\right ) ,
\end{equation*}
while the tensor associated to the Jordan structure becomes:
\begin{eqnarray}
  \mathcal{R}&=&2\frac \partial {\partial u}\otimes_s \left ( x\frac \partial{\partial x}+y\frac \partial{\partial y}+z\frac \partial{\partial z}\right
  )\cr&+&2u \left ( \frac\partial{\partial u}\otimes_s\frac\partial{\partial u}+ \frac\partial{\partial
      x}\otimes_s\frac\partial{\partial x}+  \frac\partial{\partial y}\otimes_s\frac\partial{\partial
      y}+  \frac\partial{\partial z}\otimes_s\frac\partial{\partial z} \right ).\nonumber
\end{eqnarray}
It is immediately  seen that $\mathcal{R}$ is invariant under the action of the
vector fields
provided by the linear Hamiltonian functions with respect to the Poisson
tensor $\Lambda$. We can even consider the non-local product, for instance we get
$$
 \hat Z\star \hat Y=-i\hat X ,\quad \hat X*\hat Y=i\hat Z ,\quad \hat Z *\hat X=i\hat Y.
$$

It is also easy to see that the Hamiltonian vector fields associated with
linear functions provide derivations both for the point-wise product and for the
non-local product.  Thus, the associated equations of motion  do not carry a
quantum or a classical behaviour, it is the product what distinguishes the
commutative or the non-commutative nature of the space along with the locality
or non-locality of the operation. And therefore distinguishes Classical from
Quantum Mechanics. 
\end{example}

\subsection{Dynamics}

It is now possible to write equations of motion on the space of
observables. In the  Heisenberg picture it is defined as
$$
\frac d{dt}A=\frac 1{\hbar} [H,A]_-\,.
$$

By using the `geometrization', i.e. by thinking in terms of  the dual space
$\u^{*}(\Hil)$, we find  
$$
\frac d{dt}\widehat A=\frac 1{\hbar} \, \{\widehat  H, \widehat A\}\, .
$$

As there are different algebra structures on $\u^{*}(\Hil)$ it is important to
study the compatibility of differential equations with such algebra structures.
\begin{lemma}
The linear differential equations which preserve both products correspond to
the infinitesimal generators of unitary transformations.
\end{lemma}

\section{The momentum map: relating Schr\"odinger and Heisenberg pictures}

Having geometrized both the Schr\"odinger and the Heisenberg pictures of Quantum
Mechanics, we are going to show now how they are related. For more details see
\cite{Grabowski:2006,michor,spera}. 

We have already stressed that both $\Hil$ and $\mathcal{P}\Hil$  carry, among
other structures, a symplectic one. The unitary group acts on both of them and
the two actions are related by the projection map $\pi:\Hil_0\to 
\mathcal{P}\Hil$, where $\Hil_0=\Hil-\{ 0\} $. These actions are strongly symplectic and therefore
with  associated equivariant momentum maps:
$$
\xymatrix{
\Hil_0 \ar[dd]^\pi \ar[dr]^\mu & \\
& \mathfrak{u}^*(\Hil) \\
\mathcal{P}\Hil \ar[ur]_{\tilde \mu}
}
$$

It is not difficult to see that for any Hermitian operator $A$ we have
$$
\mu(\psi )(A)=\langle \psi, A\psi \rangle =\rho_\psi(A),
$$
while
$$
\tilde  \mu([\psi] )(A)=\frac{\langle \psi, A\psi \rangle}{\langle \psi ,
  \psi\rangle } =\tilde \rho_{[\psi]}(A), \quad \psi\in[\psi]\,.
$$

It is possible to rewrite these expressions in a different form, to find:
$$
\rho_\psi(A)=\Tr (A|\psi\rangle \langle \psi | ) ; \qquad \tilde
\rho_{[\psi]}(A)=\Tr \left (A \frac{|\psi \rangle \langle \psi |}
{\langle \psi , \psi\rangle }\right ),
$$
where $|\psi \rangle \langle \psi |:\Hil\to \Hil$ denotes the rank one linear
map
 defined as $|\psi \rangle \langle
\psi |:\phi\mapsto \langle \phi, \psi\rangle \psi $.

This form shows that $\tilde \rho_\psi$ can be written as a rank-one projector
with the help of the scalar product defined by the trace.

The association between the equivalence class $[\psi]$ (element of the complex
projective space) and $\tilde \rho_{[\psi]}=\frac{|\psi \rangle \langle \psi |}
{\langle \psi , \psi\rangle }$ (rank one 
projector) is clearly one-to-one and onto. This shows that the complex
projective space, along with the symplectic structure and the Riemannian
tensor, may be identified with the minimal orbit of the coadjoint action of the
unitary group in $\mathfrak{u}^*(\Hil)$, which passes through $\tilde \rho_\psi$.

The main properties of the momentum map can be collected in the following
proposition:
\begin{proposition}
  \begin{itemize}
  \item[i)] The momentum map is equivariant with respect to the action of $U(\Hil)$
    on $\Hil_0$ and the coadjoint action of $U(\Hil)$ on $\mathfrak{u}^*(\Hil)$. In
    particular, this says that the Schr\"odinger equation of motion on $\Hil$ is
    $\mu$-related with the Heisenberg equation of motion on $\mathfrak{u}(\Hil)$
    (the space $\mathfrak{u}(\Hil)$ is identified with the dual by means of the
    scalar product defined by the trace). Moreover,
\item[ii)] $\mu^*(\hat A)=f_A$, $\tilde \mu^*(\hat A)=e_A$.
\item[iii)] $\mu^*(\{ \hat A, \hat B\} )=\{ f_A, f_B\} $ and $\tilde \mu^*(\{ \hat A, \hat
  B\} )=\{ e_A, e_B\} $.
\item[iv)] $\mu^*(R(d\hat A, d\hat B))=G(\mu^*(d\hat A), \mu^*( d\hat B))$ and for
  the other mapping  
$$\tilde \mu^*(R(d\hat A, d\hat B))=G_P(\tilde \mu^*(d\hat A),
  \tilde \mu^*( d\hat B)) +e_Ae_B. $$ 
  \end{itemize}
\end{proposition}
\proof{
  Direct computation.
}

\begin{remark}
  Had we started with the Heisenberg picture, we would be able to reconstruct
  the Hilbert space description by means of the Gelfand--Naimark--Segal (GNS)
  construction \cite{Haag:1992,Emch}. This requires the choice of a state (a functional on the
  algebra of observables). When the chosen state is pure, the representation
  will be irreducible. The Hilbert space associated with a pure state would
  play exactly the same role that our Hilbert space $\Hil$ has played for the
  Schr\"odinger picture. In this case the corresponding momentum map $\mu$ would
  provide us with a symplectic realization of the Poisson manifold
  $\mathfrak{u}^*(\Hil)$ (with the Lie-Poisson structure). We recall what a
  symplectic realization is: 
\begin{definition}
  A {\bf symplectic realization} of  a Poisson manifold $(N,\{\cdot, \cdot\} )$ 
is a Poisson map $\Phi:M\to  N$, where  $(M,\omega)$ is a
symplectic manifold. When $M$ is a symplectic vector space we have a special
situation and $\Phi$ is called a {\bf classical Jordan--Schwinger map} \cite{liejordan}.
\end{definition}

\end{remark}

\begin{remark}
  We would like to emphasize that the GNS construction brings in the quantum
  theory an entirely new framework that the traditional Schr\"odinger formalism
  is lacking of. This is due to the fact that the Hilbert space on which the
  observables act as operators is not a perennial feature of the theory, nor of
  the model to be constructed, but it is dependent on the state or preparation
  of the system under consideration. In other terms, as the state is prepared
  by the observer, the corresponding Hilbert space  with the associated
  representation of the $\C^*$-algebra is `selected' by the observer.
\end{remark}

In concluding this section we notice that the geometrical version we have
presented allows us to put the Schr\"odinger and the Heisenberg pictures within an
unified framework of Hilbert spaces, actions of the unitary group and its
associated momentum maps.

The GNS construction may be given the nice geometrical description of the
construction of a symplectic realization of the Poisson manifold
$\mathfrak{u}^*(\Hil)$, which in turn can be considered as a generalization to
arbitrary dimension of the Jordan--Schwinger map.

For completeness, let us expose a little more our considerations on the space
of states:

\subsection{States: Density states}
We have seen how the momentum map $\tilde \mu$ allows us to embed the complex projective
space $\mathcal P\Hil$ on the dual of the Lie algebra $\mathfrak{u}(N)$. The
resulting elements represent the set of pure states of the quantum system.
But in many physical situations we have more general states, i.e. density
states which are convex combinations of pure states. They are
represented by a family $\rho=\{ \rho_1, \cdots, \rho_k\}$, each element satisfying 
$$
\rho_k^2=\rho_k,\quad \rho_k^+=\rho_k, \quad \mathrm{Tr}\rho_k=1,
$$
along with a probability vector, namely $\vec p=(p_1, p_2, \cdots , p_k)$ with
$\sum_jp_j=1$ and $p_j\geq 0 \quad \forall j$. Out of these we construct a density state
$\rho=\sum_jp_j\rho_j$. The evaluation of this state on some observable $A$ is
given by
\begin{equation}
  \label{eq:aver_mix}
\rho(A)=  \sum_jp_j \mathrm{Tr}\rho_j A=\mathrm{Tr}\rho A\,. 
\end{equation}

We shall call {\bf density states}  to the set $\mathcal{D}(\Hil)$ of all convex
combinations of pure states \cite{Grabowski:2005}.

As any of the elements in $\rho$  can be embedded into $\u^*(N)$, it makes perfect
sense to consider $\rho$ also as an element in the dual of the unitary algebra.
And we hence consider the geometric structure we defined on $\u^*(N)$ as the
Poisson or the Jordan brackets
\begin{align}
  \label{eq:poissonjjordan}
  \{ f_A, f_B\}(\rho) &=\sum_kp_kf_{[A,B]_-}(\psi_k) =\sum_kp_k\{ f_A, f_B\} (\psi _k) \nonumber
  \\
(f_A, f_B)(\rho)&=\sum_k p_k(f_A, f_B)(\psi_k)
\end{align}
where $f_A(\rho)=\sum_kp_kf_A(\psi)$.

As for the geometric structures on $\mathcal{D}(\Hil)$ we shall
 consider it as  a real manifold with boundary embedded into the real
vector space $\u^*(N)$. On this space the
two structures above \eqref{eq:poissonjjordan}, define a Poisson and a
Riemannian structure. The Poisson structure is degenerate. However it is also
possible to define a generalized  complex structure satisfying 
\begin{equation}
  \label{eq:almost}
  J^3=-J
\end{equation}

The boundary is a stratified manifold, corresponding to the union of symplectic
orbits of $U(N)$  of different dimensions, passing through density matrices of
not maximal rank.  For further information see
\cite{Grabowski:2005,Grabowski:2006}.

\section{Quantum mechanics on phase space}

The phase-space formulation of Quantum Mechanics has a long history. See
\cite{cirelli1,mukunda,chaturvedi,chaturvedi2,mankomarmosimoni,mankomarmoven}
for further details. As it
stands nowadays, we may identify two basic independent ideas behind this
formulation:
\begin{itemize}
\item the first one, due to Weyl, emerged from the desire to `quantize'  classical
  systems by using bounded operators (one-parameter groups of unitary operators
  instead of their infinitesimal generators which would create  domain problems
  due to their unboundness, see for instance Wintner's theorem in \cite{EMS}. By
  a clever use of the Fourier transform, Weyl \cite{weyl,Weyl:1931}was able to
  set up a rule that 
  maps a classical dynamical variable (a function on phase-space) onto a
  corresponding operator for the quantum system in a linear manner.
\item The second idea is due to Wigner \cite{wigner} who associated a phase-space
  distribution  with each quantum state. This was motivated by the statistical
  properties of the states , in the way we mentioned in the introduction.  
\end{itemize}

It was Moyal who discovered \cite{moyal}
 that the Weyl correspondence rule can be
inverted by the Wigner map and therefore that the two approaches were exactly
inverse of one another.  As a result, the quantum expectation value of an
operator can be represented in a classical-looking form as a statistical
average of the corresponding phase-space function. In this way, Quantum
Mechanics can be represented as a `statistical theory' on the classical
phase-space. It should be mentioned, though, that as a function on the classical
phase-space the Wigner distribution (associated with Hermitian operators) is
real but not necessarily point-wise non-negative, and thus it cannot be
interpreted as a true probability distribution. The occurrence of negative
values for the Wigner function associated with states is closely related to the
impossibility of simultaneously measuring conjugated variables (as position and
momenta). 

\subsection{Weyl systems}

We consider a phase-space defined by a symplectic vector space $(E, \omega)$, with
symplectic structure $\omega$. 
\begin{definition}
  A {\bf Weyl map} is a (strongly) continuous map from $E$ to the set of unitary
  operators on some Hilbert space $\Hil$:
  \begin{equation}
\label{eq:weyl}
W: E\to U(\Hil)\,, 
  \end{equation}
such that
$$
W(e_1)W(e_2)W^+(e_1)W^+(e_2)=e^{i\,\omega(e_1, e_2)}\,\mathbb{I}\,,
$$
for any pair of vectors $e_1, e_2\in E$. The symbol $\mathbb{I}$ stands for the
identity operator on the Hilbert space $\Hil$.
\end{definition}

A theorem by von Neumann \cite{vonNeumann} asserts that such a map exists for any finite
dimensional symplectic vector space. As a matter of fact the Hilbert space
$\Hil$ can be realized as the space of square integrable functions on any
Lagrangian subspace of $E$. If we choose a Lagrangian subspace $L$, it is
possible to `decompose'  $E$ into:
$$
E\sim L\oplus L^*=T^*L\sim L^*\oplus (L^*)^*=T^*L^*\,.
$$ 

The Lebesgue measure is a translational invariant measure on $L$ and we can
construct a specific realization of the Weyl map $W$.

We define
$$
U=W|_{L^*} \quad V=W|_L,
$$
and the action on $\mathcal{L}^2(L, d^nx)$ is then given by
$$
(V(y)\psi)(x)=\psi(x+y), \quad (U(y)\psi)(x)=e^{i\alpha(x)}\psi(x),
$$
for any $x,y\in L$, $\alpha\in L^*$ an $\psi\in \mathcal{L}^2(L, d^nx)$.

Out of these two operators $U$ and $V$ we can also recover $W$ by setting
$$
W=U\circ V.
$$
But other ways of `reconstructing' $W$ are also possible.

The strong continuity condition we put in the definition of $W$ allows us to
use Stone's theorem to obtain
$$
W(v)=e^{iR(v)}\,,\quad v\in E\,,
$$
with $R(v)$ being the infinitesimal generator of the one parameter group of
unitary transformations $W(tv)$, for $t\in \R$. We have
$R(v_1+v_2)=R(v_1)+R(v_2)$.  When  we select a complex structure on the space
$E$, say by a tensor $J:E\to E$ with $J^2=-\mathbb{I}$, it is possible to define
what are known as `creation' and `annihilation' operators:
$$
a(v)=\frac 1{\sqrt{2}}(R(v)+iR(Jv)) \quad a^+(v)=\frac 1{\sqrt{2}}(R(v)-iR(Jv)).
$$

This complex structure allows to define a `Hermitian inner product'  on $E$
by setting 
$$
\langle v_1, v_2\rangle =\omega(v_1, Jv_2)-i\,\omega(v_1, v_2),
$$
where we must choose $J$ in such a way that 
$$
g(v_1, v_2)=\omega(Jv_1, v_2)
$$
defines an Euclidean inner product on $E$.

By selecting a decomposition of $E$ into $L\oplus L^*$, we may write $W$ in a
explicit form. Indeed, if we take $(x, \alpha)\in L\oplus L^*$ we set
$$
W(x, \alpha)=\exp \left ( \frac i\hbar (x\hat p+\alpha \hat q)\right ).
$$ 
Where $\hat p$ and $\hat q$ are the infinitesimal generators, as from Stone's
theorem, associated with vectors $(0,1)$ and $(1,0)$ in $L\oplus L^*$ respectively.

It is now possible to associate an operator with any function $g$ on $E$
admitting a Fourier transform $\widetilde g$. Consider thus one such function $g$
admitting as a Fourier transform
$$
g(p,q)=\frac 1{(2\pi)^n}\int d^nx\,d^n\alpha\, \widetilde g(x, \alpha)e^{i(xp+\alpha q)}.
$$
We can associate to $g$ the operator
$$
W(g)=\frac 1{(2\pi)^n} \int d^nx\,d^n\alpha\, \widetilde g(x, \alpha)\,\exp \left (
  \frac i\hbar (x\hat p+\alpha \hat q)\right ),
$$
i.e. we have replaced the Weyl exponential $e^{i(xp+\alpha  q)}$ with the
corresponding exponential Weyl operator.

This association defines a unitary isomorphism between Hilbert--Schmidt operators on
$L$ and square integrable functions on $L\oplus L^*$.

\begin{remark}
To consider the transformation properties under the symplectic linear group,
it is often more convenient to use the symplectic Fourier transform, written
as

$$
g(q,p)=\frac 1{(2\pi)^n}\int d^nx\, d^n\alpha\,  \widetilde g(x, \alpha)\,\exp \left ( \frac i\hbar (\alpha
  q-xp)\right ).
$$

The Weyl map can be then given three equivalent expressions:
\begin{itemize}
\item $W_1(x, \alpha)=\exp \left ( \frac i\hbar (\alpha \hat q-x \hat p)\right )$
\item $W_2(x, \alpha)=\exp \left ( \frac i\hbar (\alpha \hat q)\right )
\exp \left (- \frac {i}\hbar
    (x \hat p)\right )$
\item $W_3(x, \alpha)=\exp \left (- \frac {i}\hbar(x \hat p)\right )\exp \left ( \frac i\hbar (\alpha \hat q)\right )$
\end{itemize}

They are related by
\begin{multline*}
\exp \left ( \frac i\hbar (\alpha \hat q-x \hat p)\right )=\exp \left (- \frac
  i{2\hbar}(\alpha(x)\right )\exp \left ( -\frac i\hbar(x \hat p)\right ) \exp \left (
  \frac i\hbar(\alpha   \hat q)\right )= \\\exp \left (- \frac
  i{2\hbar}(\alpha(x)\right ) \exp \left (  \frac i\hbar(\alpha   \hat q)\right ) \exp \left (-
  \frac i\hbar(x   \hat p)\right )
\end{multline*}

 These relations follow from recalling that $e^{\hat A+\hat B}=e^{\hat
   A}e^{\hat B}e^{-\frac 12 [\hat A,   \hat B]}$ whenever $\hat A$ and $\hat B$
 commute with $[\hat A, \hat B]$. Correspondingly we have
$$
W_1(g)=\frac 1{(2\pi)^n}\int e^{\frac i\hbar (\alpha \hat q-x\hat p)}\,\widetilde g(x,
\alpha)\,d^nx\,d^n\alpha\,, 
$$
$$
W_2(g)=\frac 1{(2\pi)^n}\int e^{\frac i\hbar (\alpha \hat q)}\,
\widetilde g(x,\alpha)\,e^{-\frac i\hbar (x\hat p)}\,
d^nx\,d^n\alpha \,,
$$
$$
W_3(g)=\frac 1{(2\pi)^n}\int e^{\frac {-i}\hbar (x\hat p)}\,\widetilde g(x,
\alpha)\,e^{\frac i\hbar (\alpha \hat q)}  \,
d^nx\,d^n\alpha \,.
$$

The last two versions of the Weyl map are encountered in the theory of
pseudo-differential operators where one deals with symbols and operators, the
symbols being functions on phase-space corresponding to quantum mechanical
operators \cite{Hor:1965,Horm:1979}. 
\end{remark}

\subsection{Wigner's construction}

The second basic idea was due to Wigner. We shall give here a rather abstract
presentation of this idea.  It relies on the construction of two maps
, that we can
denote as
$$
U:E\to \mathcal{A}; \quad T:E\to \mathcal{A}^*.
$$
Thus we associate an operator in $\mathcal{A}$ and an operator in
$\mathcal{A}^*$ to any vector in $E$. We impose the condition:
$$
T(e')(U(e))=\delta(e'-e),
$$
where by $\delta(e'-e)$ we denote the Dirac distribution. We also ask that
$$
\int_Ede\, T(e)\otimes U(e)=\mathbb{I}.
$$
Hence we may construct a resolution of the identity from $\mathcal{A}$ to
$\mathcal{A}$ or from $\mathcal{A}^*$ to $\mathcal{A}^*$. Now, if we consider
an operator $A\in \mathcal{A}$, we can define a function on $E$ by setting:
$$
W_A(e)=T(e)(A)
$$
and reconstruct an operator on $\mathcal{A}$ from an element $f$ of  the set of
functions on $E$ as 
$$
\hat A_f=\int_E de\, f(e)\,U(e)
$$

By evaluating $T(e')$ on $\hat A_f$   we get:
$$
T(e')(\hat A_f)=\int_E de\, f(e)\,T(e')(U(e))=\int_Ede\, f(e)\,\delta(e'-e)=f(e').
$$
On the other hand:
$$
\int_E de\, T(e)(A)(U(e))=\mathbb{I}\,A=A
$$

Thus by constructing two maps endowed with the previous properties we are able
to build a one-to-one map from the set of operators onto the set of functions
of the space $E$. It is clear that the existence of a vector space structure on
$E$ plays no role and hence we can generalize this construction to an arbitrary
manifold.

The construction of these two maps requires some ingenuity and it is not a
trivial task. For this reason very often in the literature specific maps are
associated with the names of those who first constructed them.

Assuming that we are able to define the above maps  with the required
properties, it is simple to induce additional structures on the set of
functions of $E$. Consider for instance the following operation:
$$
T(e)(A_1\cdot A_2)=: T(e)(A_1)\star T(e)(A_2)=(f_{A_1}\star f_{A_2})(e)\,.
$$

It is quite obvious from the definition that this induces a product on the set
of functions which inherits all the properties from the operator algebra
structure. In particular, we can transfer the equation of motion by writing:
$$
i\hbar \frac d{dt}f=f_H\star f-f\star f_H\,,
$$
where $f_H$ is the function of $E$ associated to the Hamiltonian operator in
$\mathcal{A}$. 

The difference between classical and quantum mechanics may now emerge more
neatly because both theories are written in terms of the same vector space of
functions ($\mathcal{F}(E)$). The difference lies on the product we consider on
that set: the point-wise product is appropriate for classical mechanics, while this
new product $\star$ is appropriate for quantum mechanics.

We shall now consider more specifically the construction of these two maps $T$
and $U$ for the symplectic vector space $(E, \omega)$. The origin of our approach
can be traced back to Dirac (see \cite{dirac-rmp}). Let us try
 to explain it in simple terms.

For vector spaces $V$ admitting a numerable basis , we can define a set
$$
S=\mathbb{N}\times \mathbb{N},
$$
and define a map
$$
T:S\to \mathrm{Lin}(V,V),
$$
by means of a basis $\{e_j \} $in the vector space
$$
(j,k)\mapsto T(j,k)=|e_j\rangle \langle e_k|\,.
$$

Out of a linear map $A:V\to V$ we find a function on $S$ by setting
$$
f_A(j,k)=\langle e_k, Ae_j\rangle, 
$$
i.e. the function is given by the matrix elements of the operator $A$ with
respect to the basis we have chosen. Equivalently we can write:
$$
f_A(j,k)=\Tr(T(j,k)A).
$$

We can also get the operator from the function as
$$
A=\sum_{jk}f_A(j,k)\,|e_k\rangle \langle e_j|.
$$

The scalar product induced on the set of operators by the trace
$$
\langle M, N\rangle =\Tr M^+N \quad M, N\in \mathrm{Lin}(V,V),
$$
allows us to associate a dual element to $T(j,k)$. Thus we define $U(m,n)$ to
be:
$$
\Tr T^T(j,k)U(m,n)=\Tr |e_k\rangle \langle e_j, e_m\rangle \langle e_n| =\delta_{jm}\delta_{kn}.
$$

Moreover, the orthonormality of the basis $\{ |e_j\rangle \}$ implies the
orthonormality of the elements $T(j,k)$.  However, orthonormality is far less
important than the property of completeness.

To deal with functions on phase space we need eigenvectors of the position and
of the momentum operators, which we shall denote as $|q\rangle $ and $|p\rangle $,
respectively. We should mention, though, that we are using these elements with
the usual abuse of notation made by physicists, for these vectors are not
normalizable in the usual sense:
\begin{itemize}
\item They are indeed the eigenvectors of the position and momentum operators:
$$
\hat Q|q\rangle =q|q \rangle\,, \quad \hat P| p\rangle =p|p\rangle\,. 
$$
\item They form complete sets:
$$
\int_{-\infty }^\infty |q\rangle dq\langle q |  = \mathbb{I}=\int_{-\infty }^\infty |p\rangle dp \langle p|\,.
$$
\item Both sets are related to each other:
$$
\langle q,p\rangle =\frac 1{\sqrt{(2\pi \hbar)}}e^{\frac i\hbar pq}; \quad
|p\rangle =\int_{-\infty }^\infty dq |q\rangle \langle q|p\rangle $$
\item The norm of each element defines the Dirac delta distribution, i.e.
$$
\langle q, q'\rangle =\delta(q-q') \quad \langle p,p'\rangle =\delta(p-p')
$$
We can also state the property as
$$
\delta(\hat Q-q)=|q\rangle \langle q| \quad \delta(\hat P-p)=|p\rangle \langle p|
$$
\end{itemize}

Any Hermitian operator $\hat A:\Hil\to \Hil$ can be described as a function on $Q\times Q$ (the
Cartesian product of two copies of the configuration space), by assigning its integral
kernel to it, i.e. the `matrix elements' in the continuous basis $|q\rangle $,
$$
\hat A\mapsto f_A(q,q')=\langle q', \hat A q\rangle =\langle q, \hat Aq'\rangle ^*\,.
$$
To provide  a representation on a phase-space (say $T^*Q$), we can use the
mixed matrix elements $\langle q, \hat Ap\rangle $, which, regarded as a function of the
phase-space variables, also describes $\hat A$ completely. We have two
options  though (see \cite{mukunda})
$$
f_A(q,p)=\langle q, \hat A p\rangle, \quad  f'_A=\langle p, \hat A q\rangle\,. 
$$
For Hermitian operators both functions differ by a conjugation:
$$
A \text{ Hermitian } \Rightarrow f_A=(f_A')^*.
$$

By introducing 
$$
\hat T(q,p)=\frac 1{2\pi \hbar}\int dq' |q+\frac 12 q'\rangle 
\langle q-\frac 12 q'|\, e^{\frac i\hbar pq'},
$$
we find that
$$
A(q,p)=2\pi \hbar\, \Tr \hat A \hat T(q,p)\,.
$$

These operators are closely related to the Weyl operators we introduced
above. Indeed, it can be proved that
$$
\hat T(q,p)=\int_{E=L\oplus L^*}\frac{d^nx\,d^n\alpha}{(2\pi \hbar)^{2n}}\,e^{-i(xp-\alpha q)}\,e^{\frac
  i\hbar(x\hat P-\alpha \hat Q)},
$$
i.e. the operators we need for the one-to-one correspondence are the symplectic
Fourier transform of the Weyl operators $W_1(x,\alpha)$.

Summarizing, we have that for any state $\hat \rho$ of a quantum system, we can
define the function
$$
\rho(q,p)=\Tr \hat \rho \hat T(q,p),
$$
and the values of an operator $\hat A$ on that state will be written as
$$
\Tr \hat \rho \hat A=\int_Ed^nq d^np\rho (q,p)A(q,p),
$$
where the function $A(q,p)$ is associated with the operator $\hat A$ satisfying
$$
\int dpA(q,p)=2\pi \hbar \langle q, \hat Aq\rangle \quad \int dqA(q,p)=2\pi \hbar \langle p,\hat Ap\rangle \,.
$$

Associated to the observables we construct functions
$$
\hat A\mapsto f_A(p,q)=\int_{-\infty}^\infty dp'e^{\frac i\hbar qp'}\langle p+\frac {p'}2| \hat A (p-\frac
{p'}2  \rangle,
$$
and associated to the functions we recover operators by using:
$$
f_A(q,p)\mapsto \hat A=\int_E\frac{d^np\, d^nq}{(2\pi \hbar )^n}f_A(q,p) |q+\frac {q'}2\rangle \langle
q-\frac{q'}2 |\, e^{\frac i\hbar pq'}\,.
$$
This last formula can also be read as a decomposition of the operator $\hat A$
in the basis provided by $\hat T$.

\begin{remark}
  By writing
$$
|q\rangle \langle q| =\int_{-\infty}^\infty\frac {d^n\alpha }{\sqrt{(2\pi \hbar)^n}} \exp \left ( \frac i\hbar \alpha (q-\hat
  Q)\right )=\delta(q-\hat Q) 
$$

$$
|q\rangle \langle q| =\int_{-\infty}^\infty\frac {d^nx }{\sqrt{(2\pi \hbar)^n}} \exp \left ( \frac i\hbar x (p-\hat
  P)\right )=\delta(p-\hat P) 
$$
we can rewrite the expression of $\hat T$ as:
$$
\hat T(p,q)=\int_E \frac{d^nx d^n\alpha}{(2\pi \hbar )^{n}}\exp \left ( \frac i\hbar \alpha (q-\hat
  Q)+x (p-\hat P)\right )
$$

In this way, it is simple to prove the following properties:
\begin{itemize}
\item $\Tr \hat T(p,q)=1$
\item $\Tr \hat T(p_1, q_1)\hat T(p_2,q_2)=2\pi \hbar \delta(p_1-p_2)\delta (q_1-q_2)$
\item And 
\begin{multline*}
\Tr \hat T(p_1,q_1)\hat T(p_2,q_2)\hat T(p_3,q_3)=\\
4\exp \left ( \frac
    {2i}\hbar (q_1-q_3)(p_2-p_3)-(q_2-q_3)(p_1-p_3)\right )
\end{multline*}
\end{itemize}
\end{remark}

With these formulae, we can study how the algebraic structures of the set of
operators are transferred to the set of functions. In particular we can compute
the product $\hat A \hat B$. We can obtain it from the factors, or consider the
function corresponding to it as an element of the set of operators:
\begin{eqnarray}
\hat A \hat B&=&\int_{E\times E}\frac{d^np_1\,d^nq_1\,d^np_2\,d^nq_2}{(2\pi \hbar^{2n})}\,f_A(p_1,
q_1)\,f_B(p_2,q_2)\,\hat T(p_1,q_1)\,\hat T(p_2,q_2)\cr&=&\int_E\frac{d^np\,d^nq}{(2\pi
  \hbar^{n})}\,f(p_,q)\,\hat T(p,q),\nonumber
\end{eqnarray}
where $f(p,q)=\Tr \hat A \hat B\hat T(p,q)$. 

It is possible to prove that it is possible to write the function $f(p,q)$ in
terms of the functions $f_A$ and $f_B$ and an operation defined by
bi-differential operators:
$$
f(p,q)=f_A(p,q)\exp \left [\frac  \hbar {2i} \left ( \stackrel{\gets}{\frac {\partial}{\partial
        q}}\stackrel{\to}{\frac {\partial}{\partial p}}-\stackrel{\gets}{\frac {\partial}{\partial
        p}}\stackrel{\to}{\frac {\partial} {\partial q}} \right ) \right ]f_B(p,q)\,. 
$$

This expression is written normally using the $\star$-symbol for the product:
$$
f(p,q)=f_A(p,q)\star f_B(p,q)
$$
where 
$$
\star=\exp \left [\frac  \hbar {2i} \left ( \stackrel{\gets}{\frac {\partial}{\partial
        q}}\stackrel{\to}{\frac {\partial}{\partial p}}-\stackrel{\gets}{\frac {\partial}{\partial
        p}}\stackrel{\to}{\frac {\partial} {\partial q}} \right ) \right ]
$$

The set of functions endowed with this operation becomes an associative
algebra (convergence problems and formal series). We can consider also the
symmetric and the skew-symmetric parts of it:
$$
f_{[A,B]}=2i f_A(p,q)\sin
\left [\frac  \hbar {2i} \left ( \stackrel{\gets}{\frac {\partial}{\partial
        q}}\stackrel{\to}{\frac {\partial}{\partial p}}-\stackrel{\gets}{\frac {\partial}{\partial
        p}}\stackrel{\to}{\frac {\partial} {\partial q}} \right ) \right ]
f_B(p,q)
$$
$$
f_{(A,B)}=2f_A(p,q)\cos
 \left [\frac  \hbar {2i} \left ( \stackrel{\gets}{\frac {\partial}{\partial
        q}}\stackrel{\to}{\frac {\partial}{\partial p}}-\stackrel{\gets}{\frac {\partial}{\partial
        p}}\stackrel{\to}{\frac {\partial} {\partial q}} \right ) \right ]
f_B(p,q)
$$

Thus it is simple to prove that the limit defined by $\hbar\to 0$ leads to
$$
\lim_{\hbar \to 0}\frac 1{i\hbar}f_{[A,B]}=\frac{\partial f_A}{\partial p}\frac
{\partial f_B}{\partial q}-\frac{\partial 
  f_A}{\partial q}\frac {\partial f_B}{\partial p} 
$$

$$
\lim_{\hbar \to 0}\frac 12 f_{(A,B)}=(f_Af_B)(p,q)\,.
$$

Hence, we see how this formalism of Quantum Mechanics turns out to be much
better adapted to deal with the classical limit of quantum  mechanical
systems. For further details on the quantum-classical transition see
\cite{Lands:1998,landsman2,marmoscolarici,emch2,mackeycon,dirac-rmp}.

\section{Quantum dynamics on phase space}

It is now possible to write the equations of motion of a quantum dynamical
system on phase space. Consider thus a quantum system evolving on the space of
density matrices, thus defining a curve
$$
\rho(t)=\frac{|\psi(t) \rangle \langle \psi(t)|}{\langle \psi(t), \psi(t) \rangle }\,.
$$

The quadratic function corresponding to an operator $A$ in the evolution of the
state $\psi(t)$  defines a curve on the space of quadratic functions
$$
e_A(\psi(t))=\frac{\langle \psi(t), A\psi(t)\rangle}{\langle \psi(t), \psi(t)\rangle}. 
$$

By using the corresponding Wigner function, we can write:
$$
e_A(\psi(t))=\int_{-\infty }^\infty \frac{d^np d^nq}{(2\pi \hbar)^n}f_A(p,q)W(p,q;t).
$$
Here we denote by $W(p,q;t)$ the Wigner function corresponding to an arbitrary
time $t$. This raises the question on the definition of these
`time-dependent' Wigner functions. We can summarize their properties as
follows:
\begin{itemize}
\item $W(p,q;t)=\int_{-\infty }^{\infty}\frac{d^nx}{(2\pi\hbar)^n}e^{\frac i\hbar px}\psi\left ( q-\frac
    x2;t \right ) \psi^*\left (q+\frac x2;t \right )$ This can also  be written as
  $=\int_{-\infty }^{\infty}\frac{d^n\alpha
  }{(2\pi\hbar)^n}e^{\frac i\hbar q\alpha }\phi \left ( p+\frac
    \alpha 2;t \right ) \phi^*\left (p-\frac \alpha 2;t \right )$.
\item We have
\begin{multline*}
W(p,q;t)=2\pi \hbar e^{\frac \hbar{2i}\frac{\partial^2}{\partial p \partial q}}\Tr(\rho(t)\delta(q-\hat Q)\delta
  (p-\hat P)=\\
\sqrt{2\pi \hbar}e^{\frac \hbar{2i}\frac{\partial^2}{\partial p \partial q}}\left [ e^{\frac i\hbar
      pq}\phi(p;t)\psi^*(q;t)\right ]
\end{multline*}

\item And 
$$|W(p,q;t)|^2\leq \int_{-\infty}^\infty \frac{d^nx}{(2\pi \hbar)^n}|\psi\left (q-\frac x2\right
  )|^2\int_{-\infty}^\infty \frac{d^nx'}{(2\pi \hbar)^n}|\psi\left (q+\frac {x'}2\right )|^2=\frac
  1{(\pi \hbar)^2}$$ 
This inequality captures the uncertainty relations. 
\end{itemize}

The equations of motion for the Wigner function follow from von Neumann
equation on states
\begin{equation}
  \label{eq:vonneumann}
  \frac d{dt}\rho(t)=\frac i\hbar [H,\rho(t)].
\end{equation}

In Weyl--Wigner representation we find thus:
$$
\frac d{dt}W(p,q;t)=\frac 2\hbar H(p,q)\sin \frac \hbar 2 
\left ( \stackrel{\gets}{\frac {\partial}{\partial
        q}}\stackrel{\to}{\frac {\partial}{\partial p}}-\stackrel{\gets}{\frac {\partial}{\partial
        p}}\stackrel{\to}{\frac {\partial} {\partial q}} \right )
W(p,q;t) 
$$
where $H(p,q)$ is the Weyl transform of the Hamiltonian operator.

Then if we write the `inner derivation' associated with this Hamiltonian
`function' we obtain:
$$
D(p,q)=\frac {2i}\hbar H(p,q) \sin 
\left [\frac  \hbar i \left ( \stackrel{\gets}{\frac {\partial}{\partial
        q}}\stackrel{\to}{\frac {\partial}{\partial p}}-\stackrel{\gets}{\frac {\partial}{\partial
        p}}\stackrel{\to}{\frac {\partial} {\partial q}} \right ) \right ]
$$
We can thus write:
$$
\frac d{dt}W(p,q;t)=-iD(p,q)W(p,q;t)
$$

A formal solution for this equation can always be written in the exponential
form
$$
W(p,q;t)=e^{-iD(p,q)(t-t_0)}W(p,q;t_0)
$$

\begin{example}
  Consider for instance a Hamiltonian of mechanical type, i.e.
$$
H(p,q)=\frac{p^2}{2m}+V(q)
$$

Th equation  above becomes
\begin{multline*}
\left (\frac \partial{\partial t}+\frac pm \frac \partial {\partial q}-\frac{dV}{dq}\frac \partial{\partial p}\right
)W(p,q;t)= \\
\sum_k\frac{(-1)^k}{(2k+1)!} \left ( \frac \hbar 2\right
)^{2k}\frac{d^{2k+1}V(q)}{dq^{2k+1}} \frac{\partial^{2k+1}}{\partial p^{2k+1}}W(p,q;t)
\end{multline*}

Thus if we consider the classical limit by considering $\hbar\to 0$ we obtain that the
limit of the von Neumann equation written in terms of the Wigner function
becomes:
$$
\left (\frac \partial{\partial t}+\frac pm \frac \partial {\partial q}-\frac{dV}{dq}\frac \partial{\partial p}\right
)W(p,q;t)= 0;
$$
i.e. the classical Hamilton equations.

\end{example}

\section{Alternative Hamiltonian descriptions}

Let us recall very briefly what a classical bi-Hamiltonian system is. For more
details the interested reader is addressed to \cite{bihamilto,marmo2}.
 On the
space of functions (observables) $\mathcal{F}(E)$, a dynamical system $\Gamma$ is
bi-Hamiltonian if there exist two Poisson brackets and two Hamiltonian
functions such that the corresponding Hamilton vector fields coincide with $\Gamma$,
i..e
$$
\frac d{dt}f=\{ H_1,f\}_1 =\{ H_2,f\}_2 
$$
In terms of Poisson bi-vector fields we have
$$
\Lambda_1(dH_1)=\Gamma =\Lambda_2(dH_2)
$$

When besides this $\lambda_1\Lambda_1+\lambda_2\Lambda_2$ is a skew-symmetric tensor defining a new
Poisson bracket, the two Poisson structures are said to be {\bf compatible}.

Going over to Quantum Mechanics, it seems natural to try to imitate the same
definition, taking into account that this time one has, in addition to the
Poisson structure $\Lambda$, a Riemannian tensor $G$. We define thus:
\begin{definition}
  Consider two K\"ahler structures $(G_1,\Lambda_1, J_1)$ and $(G_2, \Lambda_2, J_2)$. A
  dynamical system $\Gamma$ is said to be {\bf bi-K\"ahlerian} if it preserves all the
  tensors: 
$$
L_\Gamma G_1=L_\Gamma G_2=L_\Gamma \Lambda_1=L_\Gamma \Lambda_2=0
$$
This already implies that $L_\Gamma J_1=L_\Gamma J_2=0$.

It is not difficult to prove that these two admissible (i.e. invariant)
Hermitian tensors are {\bf compatible} if they give rise to the same complex
structure, i.e. $J_1=J=J_2$
\end{definition}

From these two inner products on the Hilbert space, we find that the
`row-by-column' product (i.e. the corresponding coordinate expression) will
change by means of the insertion of a positive matrix $K$:
\begin{align*}
(x_1^*, \cdots ,x_n^*)(x_1, \cdots, x_n)^T=&\sum_j x_j^*x_j \mapsto (x_1^*, \cdots ,x_n^*)\cdot_K(x_1, \cdots,
x_n)^T=\\
&\sum_j x_j^*K^{jm}x_m  
\end{align*}

Thus on the space of matrices (operators) the induced alternative product
becomes:
$$
AB\mapsto A\cdot_KB=AKB\,.
$$

This implies that the Heisenberg picture we get two alternative descriptions
of the dynamics:
$$
\frac i\hbar \frac d{dt}A=[A, H_1]=[A, H_K]_K.
$$
Thus:
$$
AH_1-H_1A=AKH_K-H_KKA , \quad \forall A.
$$
Thus we obtain an obvious solution
$$
H_1=KH_K\,.
$$

For it to be admissible we need
$$
[H_1,K]=0\,.
$$
For general considerations, see \cite{bihamilto}.
From here it is possible to carry on the analysis of bi-K\"ahlerian dynamics (quantum
dynamics) along the same lines of the classical situation.

Assuming that both Poisson structures give rise to the corresponding symplectic
structures (i.e. both are non-degenerate), we can construct the corresponding
Weyl systems and the corresponding Wigner--Weyl formalisms. The corresponding
quantum dynamics:
$$
i\hbar \frac{d}{dt}f_A=f_H\star f_A-f_A\star f_H =f_{H_K}\star_Kf_A-f_A\star_Kf_{H_K}\,.
$$

By considering the deformed associative product in the form $A\cdot_KN=AKB$ we find 
$$
f_A\star_Kf_B=f_A\star f_K\star f_B.
$$

This implies that in the `classical limit' the symmetric bracket becomes
$f_Af_Kf_B$ while the skew-symmetric part becomes:
$$
\lim_{\hbar \to 0}\hbar^{-1}(f_A\star_k-f_B\star_Kf_A)=f_K\{f_A,f_B\} +f_AL_{X_K}f_B-f_BL_{X_K}f_A. 
$$
Thus the classical limit is not a Poisson bracket but a Jacobi bracket.

We may  also remark that linearly related alternative symplectic structures
give rise to Poisson brackets which are always compatible. Therefore, to obtain
classical limits (of alternative products on the space of operators) which are
not compatible we have to consider nonlinear transformations also at the
quantum level \cite{ercolessi}. Therefore, in the `Heisenberg cut' when we
describe a quantum 
system; it may be necessary to associate the Hilbert space structure or the
associative product structure of the space of operators with the `apparatus'
rather than with the `object'.

\section{Conclusions}

In this survey we have presented a brief description of the different tools
that Differential Geometry offers to describe quantum mechanical systems. We
show, first, the necessity of going beyond Classical Mechanics to describe
microscopic physical systems; and, at the same time, we obtain a series of
properties which the new description must provide.  We have studied the three
most common approaches to Quantum Mechanics, Schr\"odinger, Heisenberg and
Wigner-Weyl from a geometrical perspective. In the two first cases we have seen
that a K\"ahler structure and a Lie-Jordan structure arise naturally and play a
decisive role in the description of the dynamics. We proved also that the
Schr\"odinger and the Heisenberg pictures are related 
via the momentum map associated to the symplectic action of the unitary
group on the set of states of our system. In what regards the Weyl-Wigner
formalism, we provided an abstract description of its construction, and proved
why it  is the most suitable approach to study the quantum-classical
transition. Finally, in a very concise way, we discussed how the identification
of the geometric structures that we just mentioned above allows us to
generalize the concept of bi-Hamiltonian classical structures to the quantum
domain, by considering more than one K\"ahler structure (in the Schr\"odinger
representation) or different associative products on the algebra of observables
(in the Heisenberg picture).

We provide also a quite extensive list of references in order to allow the
interested reader to complete the topics presented in these lectures and extend
them if necessary.

  \end{document}